\documentclass[article]{elsarticle}
\usepackage{blindtext}
\usepackage[a4paper, total={6in, 8in}]{geometry}
\usepackage[ruled]{algorithm2e}
\usepackage{graphicx}
\usepackage{mathtools}
\usepackage{amsmath}
\usepackage{tabularx}
\usepackage{relsize}
\usepackage{relsize}
\usepackage{amssymb}
\usepackage{amsthm}
\usepackage{breqn}
\usepackage[font=small,skip=0pt]{caption}
\usepackage{setspace}
\usepackage{enumitem}
\usepackage{subfig}
\usepackage{float}
\usepackage{caption} 
\usepackage{tikz}
\usepackage{lineno,hyperref}
\usepackage{color}
\modulolinenumbers[5]
\usepackage[outdir=./]{epstopdf}
\usepackage{blindtext}
\usepackage[ruled]{algorithm2e}
\usepackage{graphicx}
\usepackage{mathtools}
\usepackage{amsmath}
\usepackage{relsize}

\usepackage{amssymb}

\begin{document}

\begin{frontmatter}

\title{Integration of Data Driven Technologies in Smart Grids for Resilient and Sustainable Smart Cities: A Comprehensive Review}

\author[1]{Mansoor Ali}\corref{cor1}
\ead{mansoor.ali.1@etsmtl.net}
\author[1]{Yazeed Ghadi}
\author[1]{Ijaz Ahmed}
\author[1]{Faisal Naeem}
\author[1]{Nadir Adam}
\author[1]{Georges Kaddoum}
\author[2]{Noor ul Huda}
\author[2]{Muhammad Adnan}
\author[3]{Muhammad Tariq}



\begin{abstract}
A modern-day society demands resilient, reliable, and smart urban infrastructure for effective and intelligent operations and deployment. However, unexpected, high-impact, and low-probability events such as earthquakes, tsunamis, tornadoes, and hurricanes make the design of such robust infrastructure more complex. As a result of such events, a power system infrastructure can be severely affected, leading to unprecedented events, such as blackouts. Nevertheless, the integration of smart grids into the existing framework of smart cities adds to their resilience. Therefore, designing a resilient and reliable power system network is an inevitable requirement of modern smart city infrastructure. 
With the deployment of the Internet of Things (IoT), smart cities' infrastructures have taken a transformational turn towards introducing technologies that do not only provide ease and comfort to the citizens but are also feasible in terms of sustainability and dependability. This paper presents a holistic view of a resilient and sustainable smart city architecture that utilizes IoT, big data analytics, unmanned aerial vehicles, and smart grids through intelligent integration of renewable energy resources. In addition, the impact of disasters on the power system infrastructure is investigated and different types of optimization techniques that can be used to sustain the power flow in the network during disturbances are compared and analyzed. Furthermore, a comparative review analysis of different data-driven machine learning techniques for sustainable smart cities is performed along with the discussion on open research issues and challenges.          

	     
\end{abstract}
 
\end{frontmatter}

\section{Introduction}

According to the UN report in \cite{econ2019}, the urban population of the world is expected to rise up to 68\% by the year 2050. With such a rise in population,  it is necessary to provide the citizens with standard living services as well. To curb problems like resource scarcity and pollution caused by over population, industrial and academic experts have proposed the smart city infrastructure as an optimal solution \cite{Silva2018}. 
The idea of a smart city, where diverse electronic gadgets and network architectures are combined using computational intelligence to provide collaborative integrated services and assistance is gaining prominence. Moreover, a smart city that monitors, forecasts catastrophes and creates early warnings is revolutionary in managing disaster situations and reducing deaths by producing the necessary data and analysis for the authorities. 

\begin{figure}[t]
	\centering
	\includegraphics[width=\columnwidth]{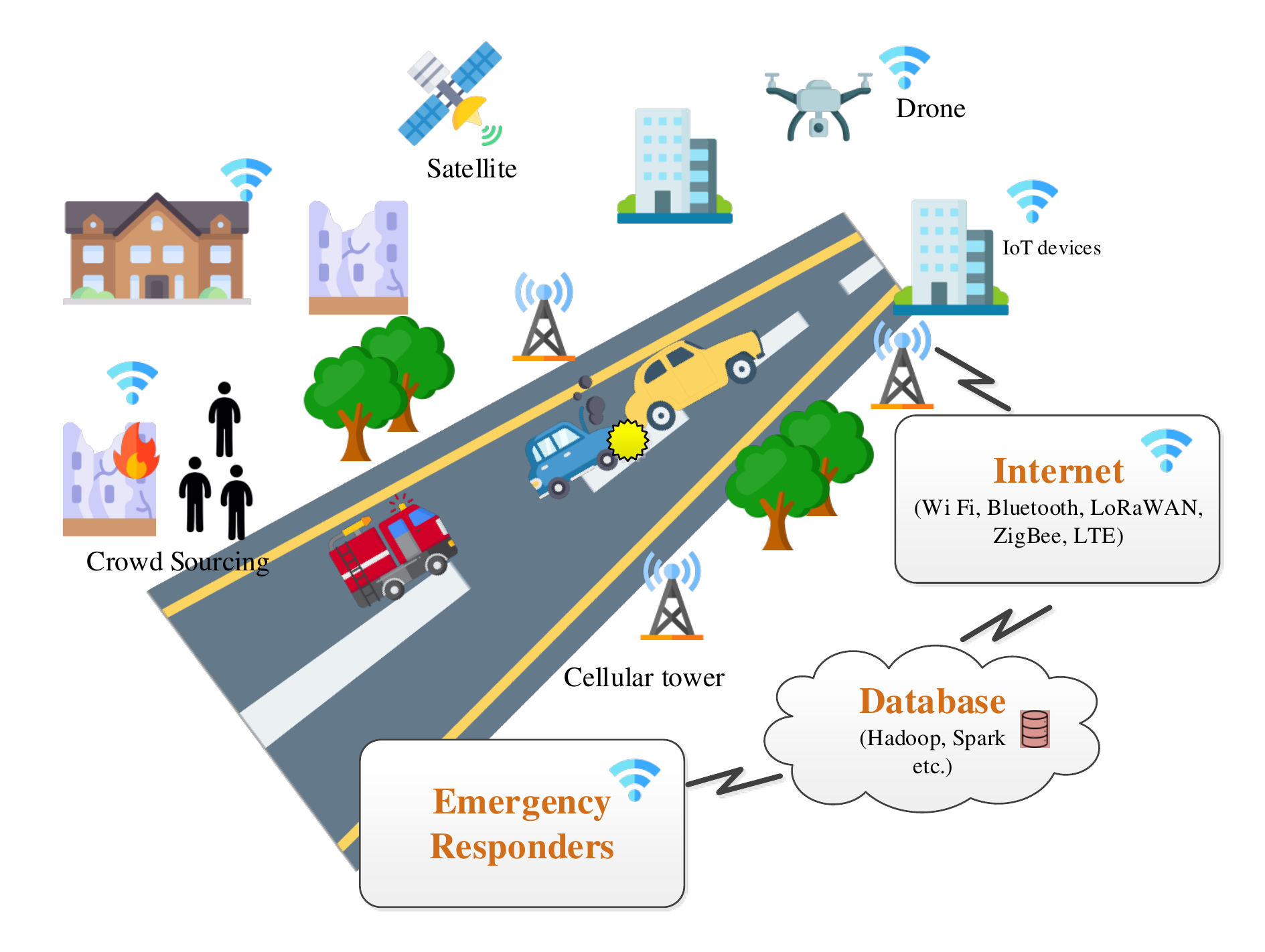}
	\caption{Role of BDA and IoT in smart cities}
	\label{IoT_smartcities}
\end{figure}

Moreover, a crucial part of any smart city is the implementation of the Internet of Things (IoT), which enable objects to connect with one another through the Internet. The ability to communicate and transmit data over the Internet using IoT devices is of immense importance in the infrastructure of smart cities \cite{Zeng2011}. However, this mutual communication among a massive number of devices generates massive volumes of data, that have to be stored and analyzed in order to produce useful information. Additionally, in the early phases of an emergency, authorities must act quickly and accurately, which is only possible if they can make quick judgments based on information that comes from a variety of sources. As it is evident from the literature, big data analytics (BDA) is the ultimate technique to deal with disparate data coming from various sources in real-time \cite{Meissner2002,Davenport2012,Mehrotra2013}.  Consequently, advanced technologies such as BDA, IoT, and computing models are playing vital roles in disaster management process automation due to their accurate and real-time results, as shown in Fig.~\ref{IoT_smartcities}. 

Moreover, the integration of IoTs and communication and networking protocols through which the network operators intelligently integrate different energy resources in a smart grid environment. These are initial steps towards a disaster-bearing smart city, as they provide the support of meshing together different data points. Nevertheless, the real-time processing of those massive volume data streams is a challenge that needs to be overcome, as it puts the lives of people at stake in the course of a disaster. This highlights the significance of BDA and the gap originated by the lack of a framework design that integrates IoT and BDA \cite{neville2016towards}.

\begin{table*}[t]

\caption{List of main acronyms.}
\centering
\def\arraystretch{1.2}
 \begin{tabular}{|p{1.3cm}|p{5.5cm}||p{1.3cm}|p{5.5cm}|} 
 \hline
Acronym & Definitions &Acronym &  Definitions                 \\ \hline
    BDA & Big data analytics& ICT & Information and communication technology\\  \hline
    RERs &  Renewable energy resources& QoS&  Quality of
service\\  \hline
    uRLLC & Ultra-reliable  and  low-latency communications&VANETs  & Vehicular ad-hoc network \\ \hline
    ML & Machine learning &
IEDs &Intelligent electronic devices\\ \hline 
   DRSC & Disaster resilient smart city &SCADA &Supervisory control And data acquisition\\ \hline
    HDFS &  Hadoop
distributed file system &EV &Electric
vehicles \\ \hline
    UAV & Unmanned aerial vehicle &DR &Demand response \\ \hline
   FANET &  Flying ad-hoc network & DRP&Demand response program \\ \hline

  IoPST &      Internet of public safety things &IBP &Incentive-based program\\ \hline
  IoE & Internet of Energy &TBRP &Time-based rate program\\ \hline
  DG & Distributed
generation & RTP& Real-time pricing \\ \hline
PMU & Phasor measurement unit &CRP & Critical peak pricing \\ \hline
AMI & Advanced metering infrastructures& DLCP& Direct load control programs  \\ \hline
API & Application programming interfaces &EDRP & Emergency demand response program\\ \hline
CAP&Capacity market programs & LOLP &Loss of load probability \\ \hline
LIHP& Low impact high probability &HILP &High impact low probability \\ \hline 
SAIFI& System average interruption frequency index &SAIDI &System average interruption duration index\\ \hline 
 LSP&Load
shedding probability& LNSP&Load not supplied probability\\ \hline 
MPC&Model predictive control&LSTM&Long short term memory\\ \hline 
 SVM&Support vector machine& BPTT&Back propagation
through time algorithm\\ \hline  
\end{tabular}

\label{table2}
\end{table*}

The development of Web 2.0 for communication; the most recent technical innovations, including social media, location-based systems, and satellites, as well as the opportunity to combine them with conventional data sources, ushers forth a new era of disaster management techniques. In contrast with conventional data collection techniques, which are both costly and complex, the integration of cutting-edge technologies has shown promise, especially in gathering and managing large volumes of data \cite{Rathore2016},\cite{Rodriguez2016}. Additionally, this new disaster management system will depend on the accessibility and fusion of information from multiple data sources.. 
In particular, utilizing social media is highly promising in this aspect, as it can be a source of dispersing information and gathering responses. In addition, social media provides smart cities with the ability to gather important data for better decision-making, as social media users contribute real-time data that can reveal more details about an occurrence \cite{Birregah2012,Goodchild2007}


Even though the role of IoTs cannot be overlooked in the holistic data collection and gathering process, however, the inclusion of IoT technologies into the structural and architectural design procedure for automated data management solutions is being highly ignored. A considerable amount of the literature is dedicated to designing disaster frameworks and models that assist emergency responders while raising awareness about the disaster \cite{Haworth2015,Houston2015,Simon2015}. In addition, traditional disaster management systems do not provide real-time and precise data during unfortunate circumstances\cite{Watson2017}. Furthermore, the task of managing and handling data coming from various sources is not to be handled by the conventional frameworks, and the pressure of producing timely and accurate results using limited resources also adds insult to the injury \cite{ortmann2011crowdsourcing}

Hence, it is high time to design an efficient framework for a smart grid environment that integrates big data and IoT with the current disaster management systems to create a resilient and reliable disaster management system for smart cities to actively mitigate the impacts of disasters \cite{Akter2017}. 

\subsection{Need of Socio-Economic Disaster Resilient Infrastructures}
Natural and man-made calamities afflicting on our world have increased at an alarming rate in past decades. According to the International Federation of Red Cross and Red Crescent (IFRC), the report on world disasters in 2018 revealed that natural disasters occurring in the past 10 years have cost financial damage of 1.6 Trillion dollars while impacting the lives of more than 2 billion people \cite{IRFC2018}. In addition, man-made activities including accidents, nuclear activities, terrorist attacks, etc., are only adding fuel to the fire, increasing the death toll even more \cite{Re2018}.In the face of such activities and calamities, disaster management measures, both preventative and reactive, have become necessary. 

The process of disaster management is vital in containing disasters' effects. In addition, disaster management processes aid in the analysis and monitoring of processes that lead to forecasting such circumstances in the future. The activities and roles of disaster management include triggering alarms with negligible false rates in addition to monitoring and analyzing real-time data. These activities are important in evaluating losses and activating damage control protocols e.g., estimating quick exit routes. These functions of damage control are generally the collective efforts of governmental and non-governmental organizations \cite{Hristidis2010}.

Although there have been certain developments in disaster management processes, the traditional path is becoming obsolete and inadequate to manage disparate resources. Cities are growing exponentially raising the need for efficient management of the massive data size pertaining to multiple resources during any crisis \cite{Baham2017}. Furthermore, the lack of adequate resources is a hindrance to the efficient processing of data arriving from several nodes \cite{Kapucu2006}. These limitations call for adopting modern disaster management measures instead of conventional methods. 

The availability of heterogeneous data sources can, however, assist disaster management systems in mitigating and containing disaster effects. In fact, obtaining data from multiple sources will make the decision-making process more in-depth and insightful. Information collection methods may consist of IoT-based sensors, satellites, as well as social media using proper social engineering. 
The developed techniques may require an organized utilization approach, however, the vitality of it cannot be ignored especially in a smart city. However, the in-charge authorities must act quickly and accurately, which is only possible if they quickly assess the credibility of the information that they receive.

Furthermore, one of the most important focus points in resilient cities is the availability of different energy resources that should be intelligently synchronized in order to keep all data sources running smoothly. Data sources in a disaster-resilient smart city ought to be able to provide information notwithstanding infrastructure problems and power outages. Backup energy-saving techniques using an intelligent integration of various renewable energy resources (RERs) and other forms of communication channels must be ensured. However, although backups, cloud-based storage systems, and remote computing can be possible solutions, the integration of smart grids to the existing framework of smart cities using computational intelligence is considered as the optimum solution to ensure their resiliency.

Using computational intelligence, a safe and resilient power grid infrastructure is considered to be one of the most important foundations for modern-day society's socioeconomic development. It is observed that due to the non-resilient nature of a power network, many countries have faced significant revenue losses. For instance, a 7-10 billion dollars loss was estimated when a severe blackout occurred in the North American power grid due to the non-resiliency of the system against high impact and low probability events \cite{council2004economic}. Moreover, 129 lines and more than 2000 substations were disconnected in China back in 2008 due to snowstorms. This disaster resulted in a complete power shutdown for more than 14 million households. Furthermore, in 2011, an earthquake in Japan caused a power outage for almost 9 days. Similarly, in 2012, Hurricane Sandy caused a power outage for millions of households on the east coast of the United States. Finally, in 2016, a tornado damaged multilevel transmission and distribution lines in China. After investigation, it was found that all these power outages were mostly due to aging and outdated electrical components \cite{schneider2016evaluating,bie2017battling}. These prolonged outages and blackouts indicate that the existing power system is not designed to handle severe and large-scale disturbances, and thus, there is a need to make a smart grid more efficient and reliable by utilizing advanced computational intelligence techniques in order to transform and developed a strong architecture of smart cities that should be resilient against such catastrophic events.

\begin{figure}[t]
	\centering
	\includegraphics[width=\columnwidth]{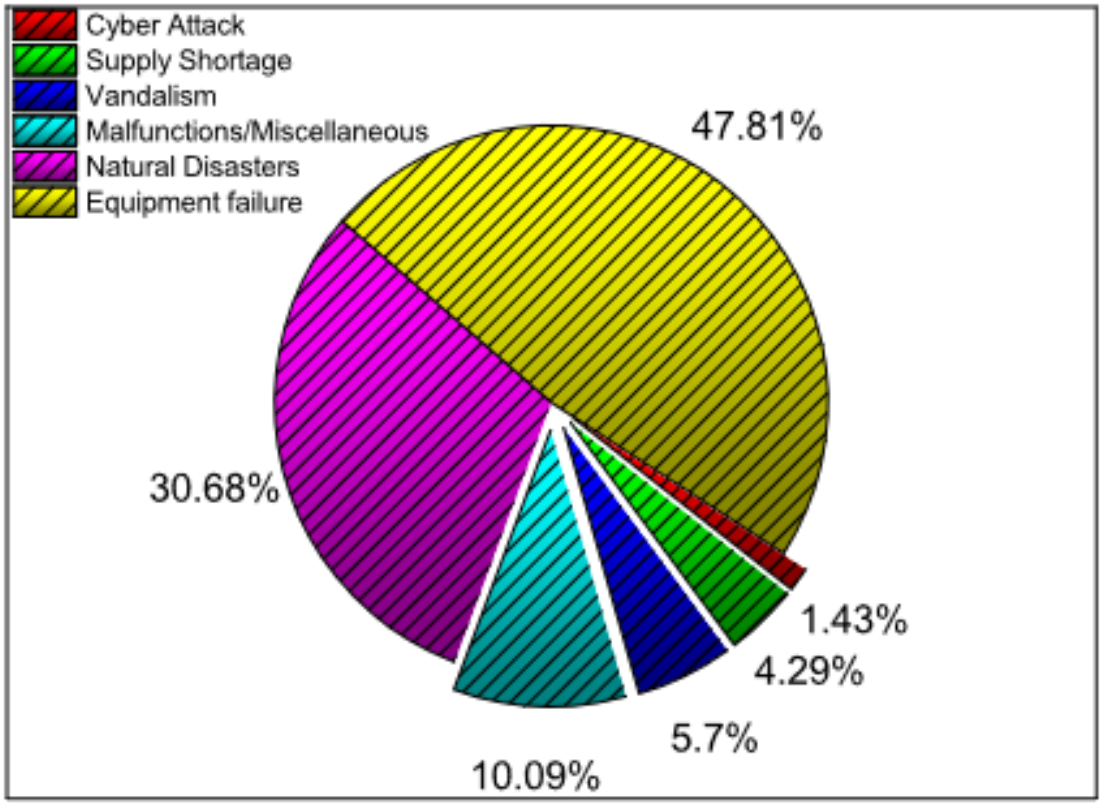}
	\vspace{0.05 cm}
	\caption{Power outages occurs due to various threats \cite{shahzad2020resilience}.}
	\label{Power_outages}
\end{figure}

Although a  power system may use advanced electronic devices, components' failure can still occur in the system due to its stochastic nature. In order to withstand stochastic component failures, system networks are mostly designed using the standard N-1 contingency principles. However, sustained outages due to extreme weather conditions, natural disasters, man-made attacks, as well as cyber attacks arise new problems in designing a better and more resilient power infrastructure. For instance, in 2013 in California, several transformers in smart grids were damaged due to gunfire by a terrorist group \cite{bie2017battling}. This shows that not only cyber but physical attacks and unexpected threats can make the designing of a resilient system more difficult and challenging. The power outage percentages due to various scenarios including natural disasters and human errors are shown in Fig.~\ref{Power_outages} \cite{shahzad2020resilience}. 

In smart cities, there are a lot of sectors ranging from smart health to power systems. In this paper, we will focus more on power systems and how they can be made resilient against disasters. In order to give a broader perspective, we will first describe the importance of BDA when it comes to designing resilient smart city infrastructure. Then, we briefly explain the role of smart drones in smart cities i.e., the use of drones for data collection and communication from IoT devices for BDA,  monitoring and management of disaster-affected areas, and ensuring the safety of localities by timely informing the concerned authorities about affected areas. Afterward, we present and discuss different parts of power systems that will be incorporated in futuristic smart cities and how to utilize BDA and IoT for disaster management and energy utilization. Then we visualize the effects of contingencies in the power system and analyze performance matrices for resiliency and reliability. Furthermore, we will provide a review analysis of the performance of different machine learning (ML) algorithms based on the refined data collected by smart drones in contingencies. This data is vital to propose a framework that is resilient to unexpected events.

This paper is organized as follows: Introduction to smart cities and the need for resilient infrastructure is described in Section 1. In Section 2, the role of BDA and IoT architecture to manage the data is presented. The information regarding smart drones and their application in various sectors is explained in Section 3. Section 4 presents power system infrastructure in smart cities, followed by Section 5, which includes resiliency and reliability performance matrices,  while resilient strategies are discussed in Section 6 and comparative review analysis is presented in Section 7. Research challenges are summarised in Section 8 and Section 9 concludes the paper.

 \section{Importance of BDA and IoT for Resilient Smart Cities}

The term big data has been used extensively in the literature to describe both its enormous size and the in-depth processing it needs in order to extract meaningful information \cite{Wang2018}. Under certain circumstances, the obtained information is highly insightful and can be used to make well-informed decisions. Although BDA’s assistance in managing the data in a uniform manner at such a large scale is highly commendable, nonetheless, issues like privacy, data breaches, and data sharing among various entities still need to be dealt with \cite{Crowley2013}.

It has been stressed quite a bit in the literature the need for precise and careful information gathering during the early phases of a disaster \cite{ortmann2011crowdsourcing}, \cite{Alamdar2016,Albuquerque2015}, which can only be achieved in practice if the information is gathered from many disparate sources such as databases and surveys. Moreover, numerous additional prospective data sources must be identified, one of which could be IoT-based sensors that provide a range of data that may be used to acquire the necessary details \cite{Poslad2015}.
Despite the fact that \cite{Shah2019}'s architecture incorporated IoT as well as BDA for transforming smart grids into smart cities, it had a few flaws, including data privacy issues and the ethical dilemma of collecting the personal information of citizens. In addition to that, it might be challenging to engage massive amounts of diverse data and extract appropriate findings in a short amount of time for emergency actions. Even with sophisticated technologies, the processing alone is time-consuming and requires complicated steps.  Furthermore, unstructured data might exacerbate the issue by necessitating various filtration techniques depending on the format which is a constraint to the real-time response.

Some recent developments to fill this gap focused on designing and forecasting danger signs and warnings but they did not account for managing the disaster effects \cite{draheim2004schema}\cite{societal2019}. In addition, numerous methods of early disaster forecasting have been established in the past including IMIS by Germany, but they lack adequate support for disaster management and do not utilize the latest technologies, which hinders them from managing big data \cite{shah2019rising}. On the other hand, an IoT-based framework for disaster-resilient smart cities that incorporates the big data framework was proposed in \cite{Shah2019}. The proposed architecture consisted of four layers for tasks like data gathering and transmission that dealt with identifying the data hubs and sources as well as means to gather and transmit the data in a complete IoT-based networked layer. These two layers were followed by the collection and management of data which made use of BDA for collecting the transmitted data, analyzing it and taking meaningful decisions based on it. In regard to big data, several other studies have been proposed regarding the implementation of IoT in smart cities. These range from healthcare systems based on big data analytics, sourcing data from IoT sensors \cite{Babar2018}, weather analysis and forecast \cite{onal2017weather}, traffic management \cite{Rathore2018}, and urban planning \cite{Babar2017}.

\begin{figure}[H]
  \centering
  \includegraphics[width=\linewidth]{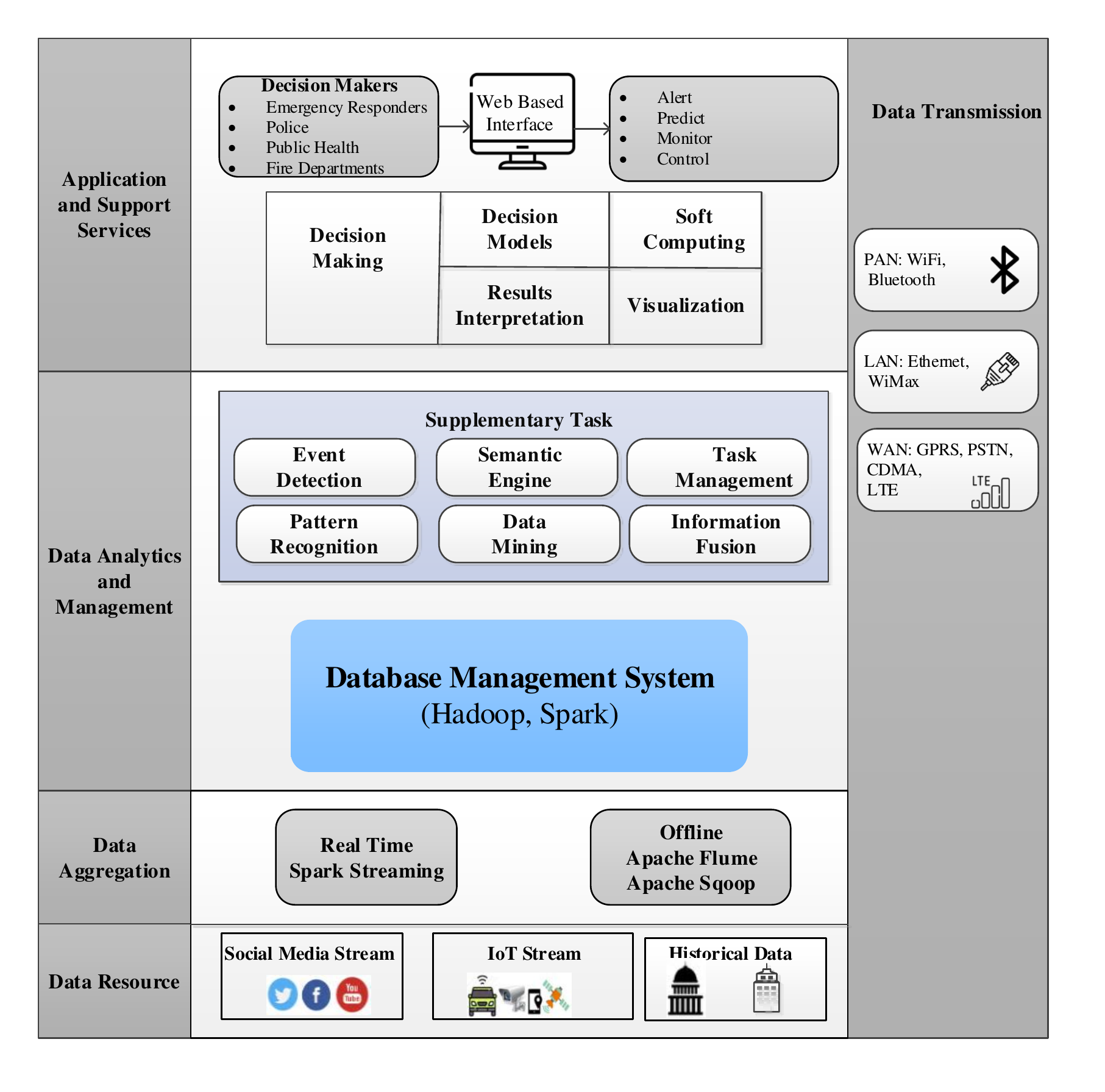}
  \vspace{-0.8 cm}
  \caption{Architecture based on BDA and IoT for disaster resilient smart cities (proposed by \cite{Shah2019}).}
  \label{fig:1}
\end{figure}

This research opens up a variety of paths for the design and evaluation of disaster resilience systems for smart cities, in particular in collaboration with IoT and BDA within smart grid environments. On these same lines, an IoT-based framework for disaster-resilient smart cities was proposed by \cite{Shah2019} incorporating the big data framework. Their proposed architecture is shown in Fig. \ref{fig:1}, and consists of four layers for tasks like data gathering and transmission that dealt with identifying the data hubs and sources as well as means to gather and transmit the data in a complete IoT-based networked layer. These two layers were followed by the collection and management of data which made use of BDA for collecting the transmitted data, analyzing it, and taking meaningful decisions based on it.

BDA-based frameworks are being used for different functionalities in smart cities, however, there is still a lack of accuracy and real-time response, and the scalability of the solutions in relation to smart cities hasn't been thoroughly explored. Therefore, there is a  motivation to address the need for a practical disaster-resilient smart city that is capable of integrating with various data hubs and power hubs based on intelligent integration of RERs to provide support to different IoT-based technologies and tools.

Hence, a part of this study will introduce and develop the concept of BDA and IoT specifically driven toward the safety and disaster resilience of smart city infrastructures. Moreover, by proposing an architectural design using state-of-the-art technologies, the solution helps in managing the heterogeneous data arriving from various hubs and their manipulation in the real-time. Similarly, the applications for disaster management require undiverted consideration given the gravity of the situation they will be used in. The process of disaster management applications may include initial forecasting, alerts and alarms, reaction and supervision, and management.

\subsection{IoT Proposed Architecture for BDA} 

In the wake of these mentioned works the architecture described by \cite{Shah2019} can be viewed as a reference for
(i)	data processing techniques establishing the relation between data and its source based on IoT in a Disaster Resilient Smart City (DRSC)
and
(ii)	 a standard for designing frameworks based on BDA and IoT for transforming smart grids to DRSC.

For the design of a disaster-resilient smart city based on IoT and BDA, several characteristics and constraints must be taken into account as mentioned below:
The architecture must be able to
\begin{itemize}
  \item Transmit the data efficiently over various communication mediums
  \item Integrate additional data sources using advanced computational intelligence in the future
  \item Store the data in coordinated and/or uncoordinated forms and be able to retrieve it effectively
  \item Scale according to the particular processing procedures
  \item Share the results and findings with the responsible stakeholder in collaborative means
\end{itemize}

Using computational intelligence to integrate smart grids, the framework architecture for smart cities can be divided into five layers: Data Resource Layer, Data Transmission Layer, Data Aggregation Layer, Data Analytics and Management Layer, and Application and Support Services. In the following, we discuss each layer briefly.

\subsubsection{Data Resource Layer}

This layer is responsible for gathering data from various sources including IoT-integrated sensors, cameras, satellite imagery, GPS, RFID tags, smartphones, social media, and data from various previous crisis databases. Depending on the sources, data can include information such as location, environment, or description of circumstances, and can also be in any format, which makes it difficult to arrange the data. Hence, this layer must process and coordinate data into meaningful information before being transferred to the next layer.

\subsubsection{Data Transmission Layer}

The transmission layer is the backbone of the architecture, as it not only connects the previous layers to the subsequent ones but also provides an effective and secure means to do so. Various means of communication and connectivity are available either wireless or wired to ensure uninterrupted transmissions on the networks \cite{zaharia2010spark}. It is also important to integrate different technologies such as 5G, Wi-Fi, Ethernet, ZigBee, and LoRA into the network.

\subsubsection{Data Aggregation Layer}

Data arrives from several different sources, so the process of analysis and decision-making can become quite complicated, which can negatively affect accuracy. The role of the data aggregation layer is to collect all the data from multiple sources and arrange them in a coordinated manner. The process can be assisted by tools like Apache and Spark Streaming. In addition, an open-source program called Apache Flume offers a distributed and dependable service for gathering, combining, and sending vast amounts of unstructured data from different sources directly to Hadoop Distributed File System (HDFS). With several recovery methods and an extendable data format for online analytical applications, it is fault resistant, reliable, and easy to use. On the other hand, Apache Sqoop is an open-source program made specifically for bulk data extraction from organized databases to HDFS. Real-time data collection from streams like Twitter is best accomplished with Spark Streaming. Using these tools in combination can help get to the desired data quickly.

\subsubsection{Data Analytics and Management Layer}

A variety of tools for data aggregation, storage, processing, querying, and analysis are included in the core layer for the data analytics and management layer. A real-time and effective solution for disaster response procedures may be created by combining several BDA frameworks. It is necessary to have an efficient and compatible storage method for streaming organized and unorganized data. The HDFS \cite{shvachko2010hadoop} is a global storage file system intended to run more efficiently on common hardware and manage massive data volumes. Its key benefit is scalability, as each machine may use local processing and storage from a central computer up to thousands of them. Hadoop's architecture is based on the master-slave configuration which is responsible for storing original information and associated metadata. In addition, Apache Spark can be used to assist Hadoop as a computation tool, due to its ability to work with APIs supporting various programming languages as well as having enough library support. 

It is essential to have a functioning event detection system for disaster management in order to detect any undesirable events. Any incident may be detected within the first few seconds of happening thanks to event detection driven by IoT-based sensor data \cite{Zhang2017}. The capacity to recognize patterns in textual or geographical data sets, which is essential for disaster management, is provided by pattern recognition mechanisms \cite{Greco2018}. For efficient information management, such as classifying, searching, and extracting unstructured data, semantic engines can be used. Furthermore, a variety of data mining techniques may be used to extract previously undiscovered patterns of observations from the given data. Then, the necessary data from diverse data sources may be integrated with the use of multi-source information fusion technology.

\subsubsection{Application and Support Services}

In order to view the results and findings of the analysis, the architecture must be equipped with a user-friendly interface that can be made public and accessible to decision-makers. The results can be presented in sophisticated graphical visualizations. Computing tools can assist the decision-making process by providing valuable insights from the results that align with the goal of the BDA-based framework of easing the decision-making process. Tools like Tableau, Kibana, Plotly, etc. can play a role in this while being integrated with conventional management systems as well.



\section{The Role of Smart Drones for Smart Cities Management}

This section outlines the functionality and importance of using IoT-based drones that are used in different applications in smart cities including data collection, energy saving, disaster management, and citizens' security.

One of the main attributes of a smart city is its ability to improve the life quality of the citizens. As IoT, disaster management, and smart energy production add to the infrastructure of the city, the use of devices that can actually connect to the Information and Communication Technology (ICT) infrastructure of the city and translate the benefits to the common citizen is the requirement of the hour. A lot of research have recently been carried out to improve the traits, features, and sustainability of smart cities \cite{lytras2018uses,bibri2017smart,azevedo2018smart,hollands2020will,mohammed2014uavs,errichiello2018leveraging}. It is evident from the cited literature that effective deployment of digital applications and information technology solutions is necessary.

\begin{figure}
  \centering
  \includegraphics[width=\linewidth]{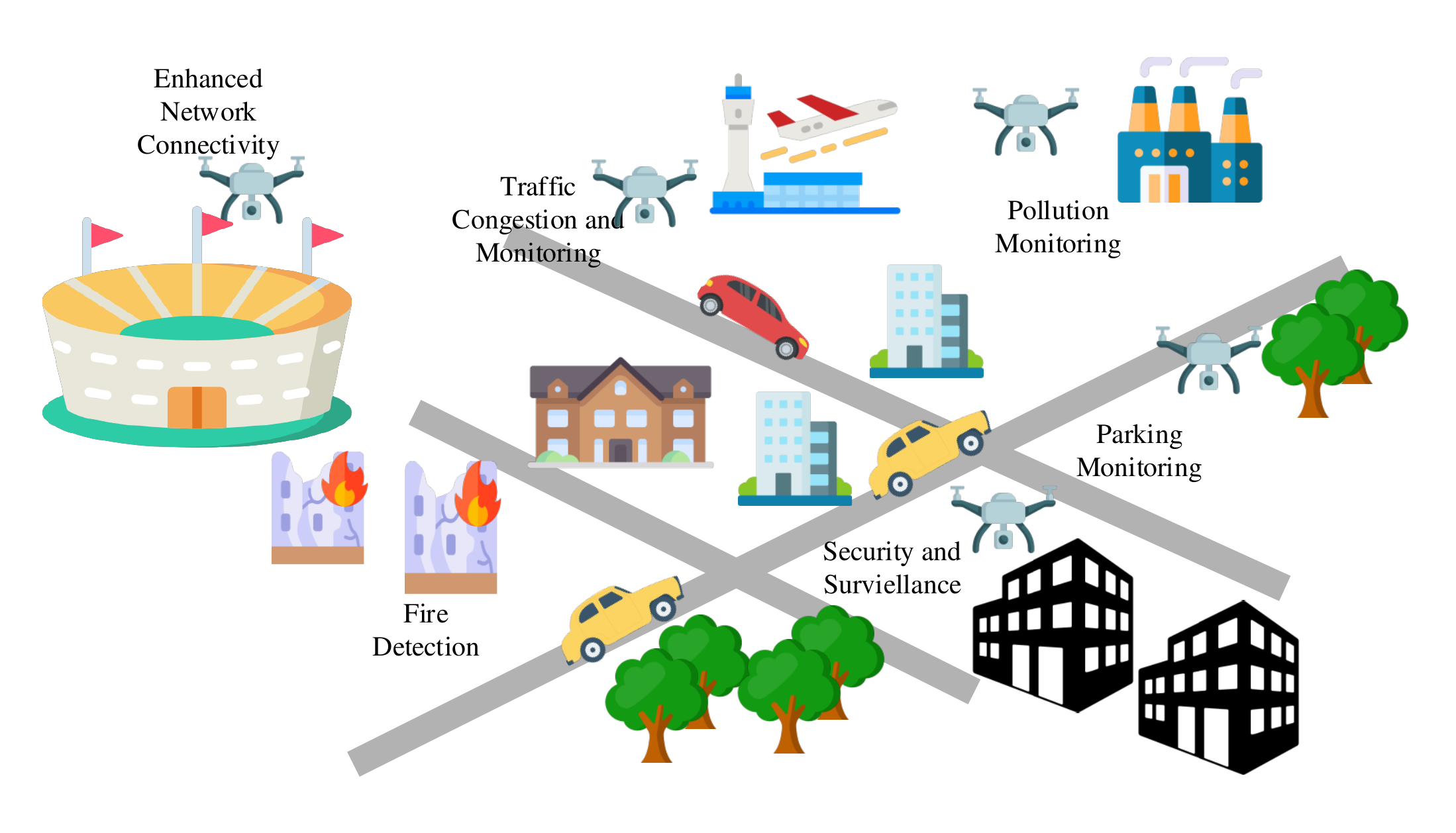}
  \vspace{0.5 mm}
  \caption{Drones used in collaboration with Smart cities.}
  \label{Drones}
\end{figure}

Unmanned aerial devices or drones are automatic flying objects that will be significantly used in smart cities for autonomous tasks like information exchange, movement, agriculture, safety and security, disaster preparedness, and environmental conservation. Unmanned vehicles and aerial devices can transmit to users the benefits of smart cities' automation, such as energy conservation, health care services, monitoring services, and waste reduction. Fig.~\ref{Drones} illustrates an overall vision for using drones within a smart city infrastructure.

In addition, drones play a crucial part in the IoT ecosystem due to their agility and line of sight (LoS) capabilities. The marvel of IoTs has introduced remarkable advances in applications, such as smart grids based on intelligent integration of RERs, cities, parking, etc. within the smart cities pertaining to the goal of IoT that everything in the IoT network is connected and communicates among themselves. The idea and design of such methods have now become indispensable to economic development and the environment in the race to conserve resources and mitigate waste. There is a need for tools and devices that can share data while coordinating with each other. However, this idea can face constraints in the form of cost-effectiveness and energy conservation as the number of connected devices increase \cite{rani2015novel,gapchup2017emerging,huang2014novel}.

\begin{figure}[h]
  \centering
  \includegraphics[width=\linewidth]{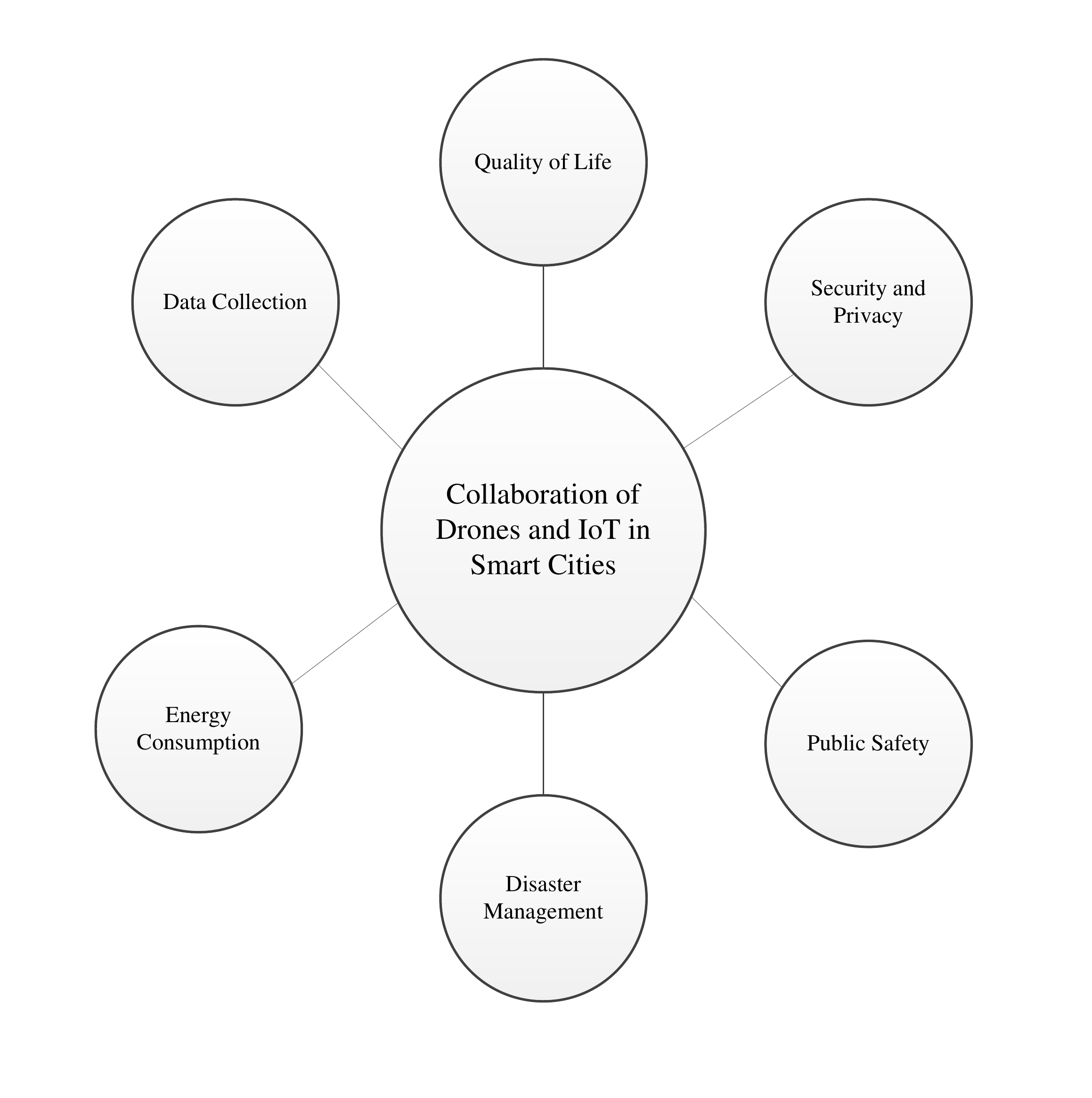}
  \vspace{-0.8 cm}
  \caption{Use cases of drones in smart cities.}
  \label{Drone}
\end{figure}

Drones are able to move autonomously in the direction of IoT devices, collect data from them, connect to them in real-time, analyze the data, and then communicate the information to other devices or collecting nodes \cite{lien2011toward}. In \cite{mozaffari2017mobile,alsamhi2018predictive}, researchers examined the effective implementation and agility of drones to gather information from ground IoT devices with minimal transmission power and improved communication stability and connection. The required IoT transmission power decreased by 45\% while the productivity increased by 28\%. Drones were also mentioned as a medium for energy-efficient research in IoT and monitoring systems in another relevant literature source \cite{sharma2017energy}. In addition, the authors of \cite{motlagh2017connection} proposed a steering strategy between several 4G networks to guarantee dependable communication of IoT devices with drones. This strategy provides increased quality of service (QoS) while using a reasonable amount of energy. A multi-clustering strategy of IoT devices was developed in \cite{mozaffari2016mobile}, in which each cluster was supported with at least one drone. Cooperative wireless networks connecting drones and smart devices are also essential to solving the routing loop issue in conventional sensor networks and extending the lifespan of sensors through load distribution. However, for an optimized drone and wireless sensor network coordination, these solutions rely on the creation of an ideal topology and routing system. The use cases for aerial drones as discussed in the subsequent section are depicted in Figure~\ref{Drone}. 

\subsection{Introduction to Smart Drones}
The introduction of drones and their possible use in smart cities was presented by \cite{Jensen201667}. In addition to discussing how drones would improve applications for smart cities, such as monitoring, supervision, and object detection, decentralized processing applications, such as gathering data and path navigation were also mentioned. Additionally, the authors in \cite{alsamhi2020convergence} explored the integration of a number of ML algorithms for effective and efficient navigation to improve connectivity, portability, and QoS.

The Flying Ad-hoc Network (FANET) system is used by a cluster of drones to communicate with one another and to share the onboard modules' collected data with other IoT devices on the ground. Drones have the capacity, portability, and responsiveness to effectively improve the QoS of 5G mobile networks. Additionally, they have been used to increase cellular coverage and bandwidth for transient events like sporting competitions, disaster relief efforts, criminal surveillance, etc. \cite{valcarce2013airborne}. In order to help disaster relief efforts when regular wireless networks are disrupted, drones have also been utilized \cite{bucaille2013rapidly}. On-board processors and sensors coupled to other devices may be part of a drone package that enables human control. It may also be designed to do tasks and navigate itself under the control of its internal systems. Additionally, it can be controlled remotely to manage agriculture and security.

Additionally, drones may be a useful tool for tracking deforestation and assessing environmental factors, such as the quality of the air and water. Furthermore, to detect and count endangered animals, trackers are utilized, which may be observed by drones. Information, connection, applications, and the sort of devices being utilized must all be converged in the design of smart drones. 

\subsection{Data collection using Drones}

The process of data collection is an elaborate and sophisticated process, pertaining to a number of parameters that need to be taken care of. The introduction of drones in collaboration with smart cities and smart grids environment has paved the way to ease in collecting data while reducing constraints.

In many smart cities, small and battery-constrained smart IoT devices are dispersed to collect information about the environment \cite{7736615}. Due to energy limitations, these IoT devices cannot send the signal across long distances \cite{5741148,7841993}. The decrease in energy consumption is a bare minimum requirement for smart cities to live up to their claims of eco-friendliness and sustainability, and the mitigation of pollutants and harmful waste is also necessary to improve the living experience \cite{7317502,7396150}. In order to collect data from minimal edge devices like environmental sensors, monitoring equipment, and other IoT devices, drones can also serve as a critical technical component in IoT device communication \cite{7207365,schaub2016drone}. In \cite{4749746}, the authors discussed how drones and IoT devices may be used collectively in a disaster reaction environment and gave the tools for automatic network restoration and event detection. 

Smart drones integrated with IoT devices can gather, store, and interpret data to carry out complicated tasks efficiently and effectively, and thus provide a solid path for transforming the smart grid environment into sustainable, efficient, and reliable smart cities. To the extent that they may interact with and react to the environment similar to the various objects in the IoT framework, drones have recently begun to symbolize a subset of the full IoT domain. Hence they may be mounted at various places, carry variable loads, and offer insights on it everywhere, anytime. Lower costs, improved connection, and the provision of high QoS are interesting features of the partnership between drones and the IoT.
The IoT apparatus onboard a drone may contain gadgets like sensors, actuators, cameras, and wireless communication tools. Thus, drones can easily gather data from IoT devices by using the technology already embedded in their weight. Drone service delivery is carried out through technologies like WiFi, LTE, and 5G.

In order to collect data from gadgets like environmental sensors, health care monitoring equipment, and other IoT devices, drones may also serve as a critical technical component in IoT device communication  \cite{dawy2016toward,hassanalieragh2015health,sathyamoorthy2015energy}. In \cite{erman2008enabling}, the authors discussed how drones and IoT devices might be used together in a disaster response environment and gave the tools for autonomous network restoration and activity recognition. 

\subsection{Disaster Management using Drones}

In addition to providing services based on ease and comfort, the idea of a smart city also encompasses efficient and effective preparations for emergency and critical circumstances by carrying out search and rescue operations in disaster-stricken areas. In this situation, applying artificial intelligence (AI) approaches for visual analytics is essential to quickly detect objects and events. Visual documentation of disaster zones can have important influences on choosing an effective disaster response. Considering this, researchers investigated the efficiency of ML to manage catastrophes in smart cities \cite{7832322}. The proposed method was applied to categorize pictures of locations affected by disasters, with great precision. Additionally, for disaster response in smart cities, the concept of smart transportation with the aid of cloud computing has been proposed. Researchers in \cite{8538356} proposed an IoT-based disaster reaction and perception system for enhancing rescuing operations. IoT-based sensors employed in this situation were able to collect data, locate injured people, and highlight hazards. The proposed system's architecture features intelligent sensors, intelligent processing, and intelligent acknowledgments.

During the circumstances of a disaster, the authors of \cite{sakhardande2016design} used different IoT device connection channels to monitor and coordinate search and rescue efforts. Additionally, for disaster management systems, the concept of smart transportation using cloud computing and VANETs has been discussed in \cite{alazawi2014smart}. Additionally, for enhancing rescue operations during a disaster, Boukerche et al. \cite{boukerche2018smart} proposed an IoT-based disaster reaction and detection system by using IoT devices to collect data, locate injured individuals, and highlight risks. The proposed system's design featured intelligent sensors, processing, reactions, and an ad hoc communication network.

The authors in \cite{aljehani2019safe}, proposed that drones integrated with IoT devices can be used to gather information by tracking and processing image data from disaster-affected areas. 3D technology can also be used to make renditions and maps of the affected premises to assist the rescue services by analyzing the maps and devising optimized paths through AI. Carefully selecting the deployed communication protocols can improve the QoS and communication levels.

Since the circumstances during a disaster are highly time-constrained, IoT-based data accumulation can help save valuable moments and accelerate search and rescue operations. For instance, the rescue team can be equipped with wearables with real-time information of the surroundings and stats.

\subsection{Public Safety using Smart Drones}

Owing to its functional capacity over distant areas, the deployment of drone technology is the most affordable and effective way to achieve the required tasks including keeping an eye on a fleeing criminal, finding and examining a disaster site, looking for a lost individual, etc. Consequently, in order to attain and deliver maximum benefits, the best characteristics of drone and IoT device coordination are efficient data collection, the ability to connect and recognize objects, accessibility, communication capability, etc. Emergency responders can communicate with command centers and other relevant parties by using wearable internet of public safety things (IoPST) devices that link smart cities with public safety. These linked gadgets can assist monitoring and relief teams in determining when and where catastrophes or crimes have occurred.

The authors in \cite{mozaffari2016unmanned} examined the use of drones as a BS that supply network connectivity to a specific region. A collaborative approach among drones and connected devices is crucial for improving public safety and minimizing the impact of catastrophes in smart cities.
The IoT's capabilities and strategies for disaster management were examined in \cite{ray2017internet}. In order to provide network connectivity between all the corresponding responders on the accident site, \cite{motlagh2017connection} used an artificial neural network (ANN) to forecast the signal strength between the drone and IoT devices which can be wearable. Finally, IoT and big data are crucial for disaster management, according to Reina et al. \cite{reina2018evolutionary}. The study concentrated on using drones as the "$0^{th}$ responders" that can reach the accident scene before the first responders to provide communication assistance to disaster victims.

\section{Power Systems in Smart Cities: A Case Study}
\label{Smart_Grids_IoT}

\begin{figure}[h]
  \centering
  \includegraphics[width=\linewidth]{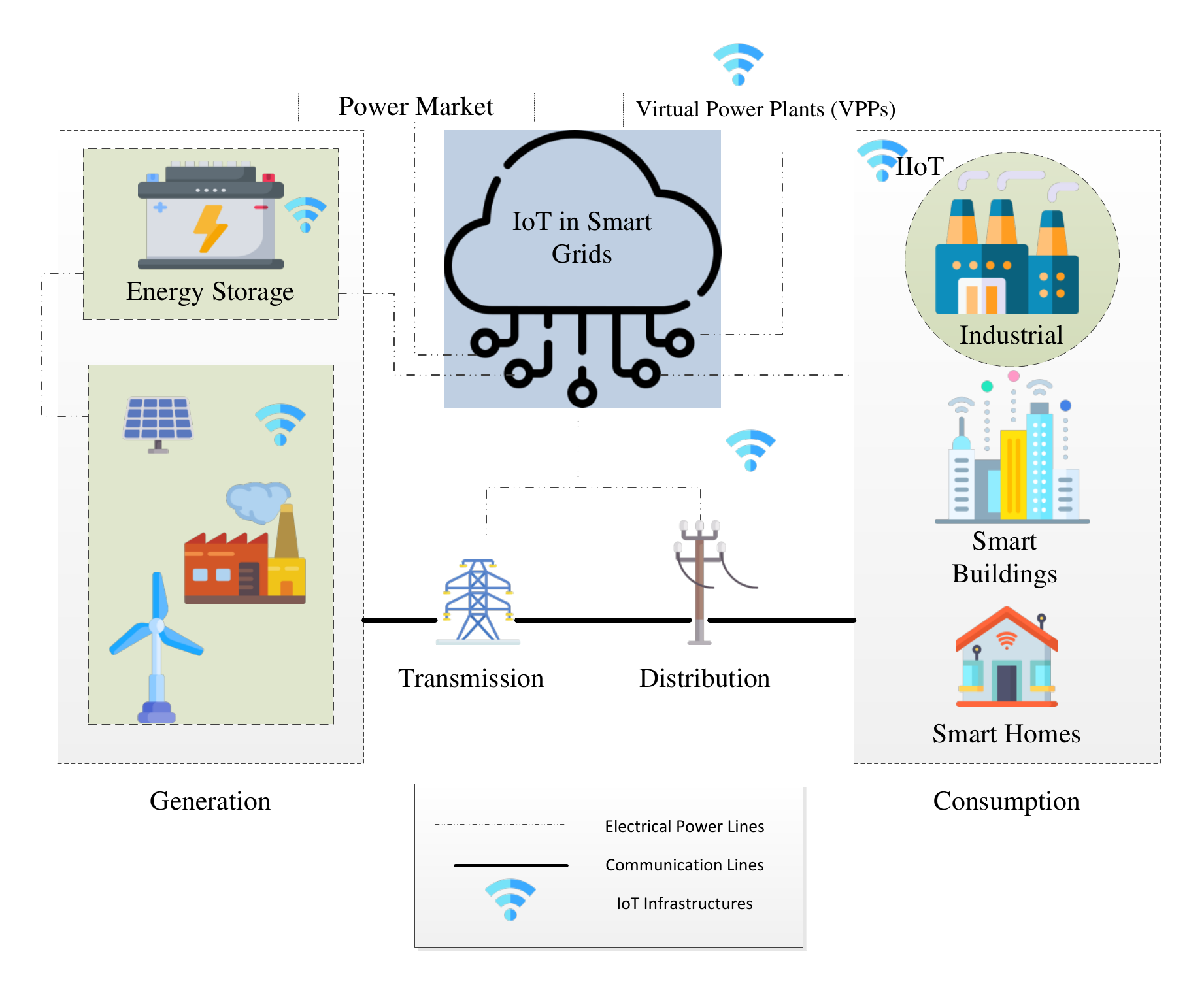}
    \vspace{-0.3 cm}
  \caption{The paradigm of IoT at smart power systems in smart cities.}
  \label{fig:flow_electrical}
\end{figure}

The transmission of energy from the producer to the consumer depends on transmission lines and distributed generation in any grid system. Moreover, any grid must have the same dynamics at the generation, transmission, and distribution levels to avoid faults. Internet of Energy (IoE) is an upgrade in the field of energy management in integration with IOTs. Smart grid 2.0 introduces the idea of utilizing the concepts of IoT i.e., communication with several devices over various protocols in real-time, into the smart grid to increase its productivity while dealing with constraints, such as the variation of energy parameters and the non-coherence of some factors \cite{buyya2016internet}\cite{cao2013energy}. Through the Internet of Energy, we hope to improve accessibility by providing access to tracking energy consumption, as well as providing power back to the grid through small-scale power generation, which is currently not feasible by conventional methods. The flow of smart electrical grids starts from generation and reaches the distribution and consumption end, as shown in Fig. \ref{fig:flow_electrical}. 

\subsection{Generation Phase}

The generation level of a power system is the base providing energy to any city, hence, it requires continuous monitoring and control using advanced computational intelligence techniques. Conventionally, locally controlled devices were deployed to regulate the management of generating resources. Due to the difficulty of handling the operating system remotely, many operations were performed by transmitting orders or instructions to a local operator \cite{shahinzadeh2019iot}. Additionally, asset management for power generation systems is becoming more advanced. Since traditional energy sources are near depletion, the need for RERs is vital. In addition, increasing the usage of electric cars will have an influence on the scheduling of power system generation in the near future. Moreover, to provide uninterrupted supply to consumers, small-scale generation in the forms of distributed generation (DGs) has been established which, in turn, requires a careful connection with the main supply to prevent the shedding of loads and fluctuations \cite{calise2021smart}.

All the above-mentioned scenarios highlight the significance of the generation levels in a power station for its stable operation. Moreover, to improve the efficiency of production, reliable operation of RERs in power systems is necessary  \cite{pal2021comprehensive}. However, the effective integration and monitoring of RERs, which can be achieved by using IoT devices, has not been extensively discussed in the literature. 

\begin{figure}[h]
  \centering
  \includegraphics[width=\linewidth]{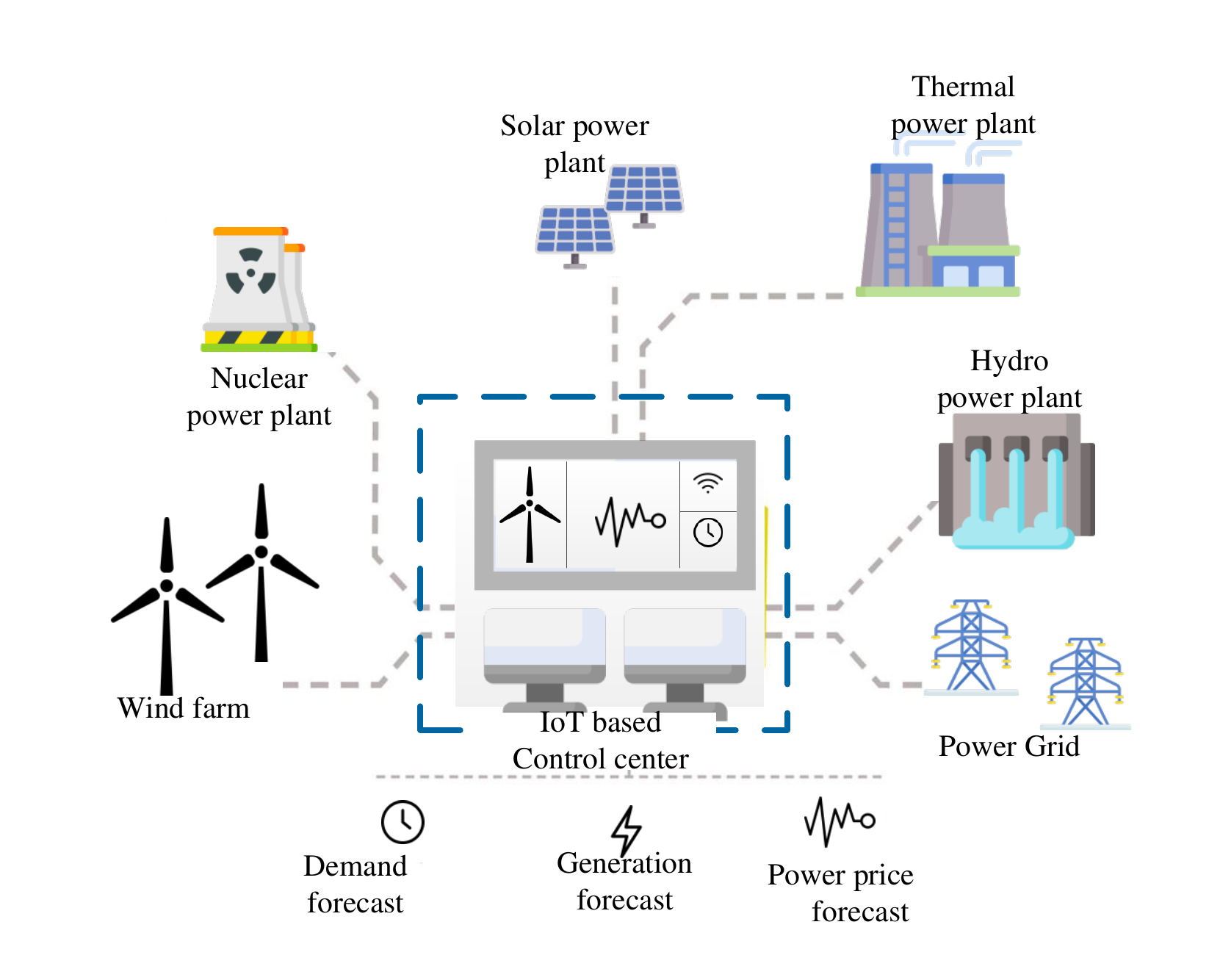}
    \vspace{- 0.3 cm}
  \caption{IoT-based real-time generation monitoring}
  \label{flow_electrical.}
\end{figure}

\subsection{Transmission Phase}

The link between the generation layer and distribution layer is the transmission level which is an indispensable fragment of power systems guaranteeing a steady supply. IoT integration with the grid at the transmission phase is crucial since IoT can help with better transmission management to avoid congestion while keeping the transmission systems secure. Intelligent electronic devices (IEDs) with IoT capabilities deployed in the transmission layer can alert the operator of changes in the lines' conditions, such as disruptions \cite{tightiz2020comprehensive}.In addition, the phasor measurement units (PMU) can be deployed along transmission lines to determine the phase and magnitude of the voltage and current at any particular point in the line, which allows the power system events to be analyzed dynamically at any moment \cite{chen2011integration}.

With standard SCADA (Supervisory Control And Data Acquisition), which reports one measurement every 2 or 4 seconds, such a quick and precise measurement is not achievable. Meanwhile, PMU and protective relays can be used to construct wide-area protection schemes \cite{Eissa2010} in which PMU measurements can go up to 120 samples per second, which increases the chances of preventing catastrophic blackouts. This leads to proactive control measures and techniques in which the operator can automatically control congestion in crowded power networks or locally congested locations, especially in crises and contingencies, thanks to the monitoring of real-time power going through the cables. As a result, the operability of the power grid can be enhanced significantly \cite{Markovic2013}.

Furthermore, natural calamities might damage overhead cables. Strong winds and heavy snowfall caused lines to gallop, which exerts an asymmetrical pulling force on the lines and may cause towers to tilt. These elements harm overhead lines, increasing the operation's danger. Additionally, the transmission system is dispersed over a huge region of land, making it challenging to maintain and control. The use of IoT can lessen the harm that such natural disasters inflict. Advanced sensing equipment mounted on the line's conductor of towers must provide the necessary data that is to be transmitted to the sync node device, then over an optical fiber network or wireless communication channels, to the central controlling center. Data collected for this purpose may include but are not limited to, sync node devices, tower deviation, temperature, humidity, wind speed, conductor acceleration, the sag of the wire, and current leakage, among others. The conductor, insulations, and towers may benefit from improved real-time monitoring thanks to these devices.

\subsection{Distribution Phase}

The installation of IoT infrastructure at all pivotal distribution network nodes is essential for a smart distribution grid to transform into a smart city. From the perspective of the consumer, installing advanced metering infrastructures (AMIs) is the first step in implementing a smart distribution grid. AMI systems consider the communications network as one of its most important components. It offers safe, reliable, and bidirectional links between servers, data gatherers, counters, clients, and beneficiary businesses.
Depending on the local environment and investment budget, this communication infrastructure may be deployed in a variety of ways. The implementation of IoT in the distribution layer results in measurable advantages such as online monitoring of consumer consumption patterns, intelligent control of energy generation and consumption, detection of issues in low-voltage transmission lines, implementation of emergency demand response programs, deployment of self-healing schemes, management of power losses, and remote monitoring and control during unplanned disasters, among others. Additionally, in order for the distribution operator to be able to conduct rigorous monitoring and supervision of the distribution grid, the acquired data from all feeders and buses must be digitalized and shared through regional ICT-based networks. The self-healing mechanism is also one of the crucial and essential plans for the futuristic distribution grid in order to increase the grid's resilience. To swiftly restore the necessary and required functionality, self-healing techniques must operate in real-time \cite{Bessis2014}.

As discussed above, an intelligent integration of RERs at the generating phase to best serve the distribution side can only be completed if a strong and resilient transmission and distribution network is designed utilizing advanced computational methods. Through this, network operators will transform a smart grid environment into a strong foundation of smart cities that have the capability to manage disaster situations.

\subsection{Microgrids}
\label{subsec:1}

Microgrids are renewable energy source-based infrastructures, that are connected to the main generation grid and are able to meet the requirements by supplying micro-sources of their own \cite{moazzami2018optimal}.
Additionally, an energy storage system can be used to store excess small-scale renewable energy production when the demand is lower. The construction of microgrids may be stand-alone or off-grid buildings, particularly for distant places or a structure that is connected to the main grid, nonetheless, they mainly work in collaboration with the main grid. Collaboration based on conventional methods, however, has resulted in low efficiency and quality. Hence, IoT-based solutions are now being proposed.
In order for the microgrid to function, the micro sources need to be functioning consistently with the main grid, however, the lack of observability makes it difficult, which can be improved using IoT. The use of IoT-based sensors on the microgrid can transmit the grid parameters to the cloud which can be accessible to all the system members. This not only improves performance but also enables multiple microgrids to be connected to each other and to the main grid in real-time.

\subsection{Smart Grid 2.0}

The first generation of smart grids paved the way for technologies like AMIs, whereas the current generation is promoting decentralization by utilizing IoT, laying the foundation for the macro-level. Due to its self-reliance, it can also provide robustness to the system by dealing with undesired obstructions. Moreover, the platform for P2P and third-party access can be done using application programming interfaces (APIs). Future grids will likely include increased microgrid deployment, automation, digitization, globalization of generating units, and energy sharing, among other aspects.

The second generation of smart grids, often known as smart grid 2.0, is a recent idea that refers to the next generation smart grid architecture \cite{cao2013energy}. There are significant differences between smart grid 2.0 and smart grid 1.0. With smart grid 2.0, the interaction between supply and demand will be implemented using sophisticated smart metering infrastructures, the proportion of energy and its congruent data will be handled by informatics, and plug-and-play capability will be in place, symbolizing the possibility of providing energy from small-scale resources (e.g., vehicle to the grid). This will enable the consumers to be able to provide power by connecting to the grid at any time without any issues from the grid dynamics.

Additionally, in smart grid 2.0, the control and monitoring of the grid equipment will be performed by the grid operator, allowing flexible control, which requires the use of BDA and other smart technologies, which in turn will improve the efficiency of the electricity market. The performance can be further improved by designing a peer to peer based interface for trading that can be accessible to all stakeholders.
 
 \subsection{Energy Saving in Smart Cities using IoT}

IoT has played a significant role in improving smart city functions through the use of disparate sensors and actuators. These functions include managing traffic systems, water supply systems, and waste management systems. Energy can be saved by using sensors such as streetlights and electric vehicles (EVs).
Parking spaces for EVs are crucial from the perspective of distribution operators, especially for vehicles with the capacity to connect to the grid i.e., vehicle-to-grid (V2G). Over the past ten years, the prevalence of EVs has increased and will continue to do so in the next years. This issue has a significant impact on how well distribution systems work. Due to the highly unpredictable nature of EV action, modeling it is also incredibly complex, nevertheless, their batteries may be recharged or discharged whenever they desire in an IoT-based system. Consequently, if the EVs can establish a connection with a cloud-based management system when they are in a certain area, the charging can be monitored.

 Additionally, control systems for parking lots can provide access to real-time control over them. In the near future, the impact of V2Gs on power grids will be a contentious topic. Because of this, operators of the electricity system would prefer to make use of IoT infrastructures than be forced to use unfavorable methods like power outages and limitations, which might result in paying damages or dissatisfied consumers.
 
Smart homes and smart buildings are considered key elements to transforming smart grid infrastructure into smart cities. In terms of quantity, residential and commercial customers make up a sizable portion of loads. Because the overall effects of each load are not insignificant, operators must pay attention to how each load behaves specifically. As a result, several cutting-edge gadgets are created to assist in lowering electricity consumption and boosting efficiency while maintaining the essential energy of a structure. Even though using IoT during building construction results in higher expenses, it also yields noticeable, measurable advantages. Utilizing IoT in houses and buildings offers creative ways to transform conventional structures and create an environment that is more effective, pleasant, ecological, and safe \cite{Stojkoska2017}.

One of the core components of smart buildings is the architecture for smart meters. The power management center has to get access to real-time electricity and gas prices in order to adjust generation and consumption appropriately. For the purpose of sending and receiving real-time signals, all energy-related devices and gadgets must be IoT-enabled. In a cloud data center, a wireless platform may successfully establish a default gateway for all devices to communicate information \cite{dey2016home}.

Moreover, the devices can be equipped with LTE and 5G technologies to enable control remotely. Similarly, the heating, ventilation and air conditioning (HVAC) system and the lighting system can be operated automatically using the data input from sensors based on the occupancy in the room. Using the combined heat and power units can drastically change the energy efficiency based on the optimum time for generating the unit. Furthermore, the appliances and devices can be employed with IoT-based protocols, such as ZigBee for automation and control. The excess consumers' power can be returned to the grid saving electricity. Moreover, the V2Gs may be charged other than the peak electricity usage hours which can be managed using sophisticated cloud-based IoT infrastructure of smart homes.
 
The traditional method for calculating consumer demands relied upon entering several parameters, which did not yield appropriate results in real-time scenarios due to the variability of parameters, thereby increasing energy costs. The use of IoT-based sensors allows the calculation of consumer demands in real-time, saving energy in smart homes.
 
The concept of demand response (DR) resulted from the development of this theory. Customers were encouraged to sign an agreement that represented a certain kind of demand response program (DRP), either voluntarily or involuntarily. DRPs can be classified as incentive-based programs (IBP) and time-based rate programs (TBR) whose characteristics include real-time pricing (RTP), time of use (TOU), and critical peak pricing (CRP) in the case of TBR \cite{Moghaddam2011}. IBP programs may include emergency demand response programs (EDRP) and direct load control programs (DLCP) as voluntary or mandatory programs, interruptible/curtailable services (I/C), capacity market programs (CAP), and market clearing programs.

\begin{figure}[H]
  \centering
  \includegraphics[width=12cm, height=10cm]{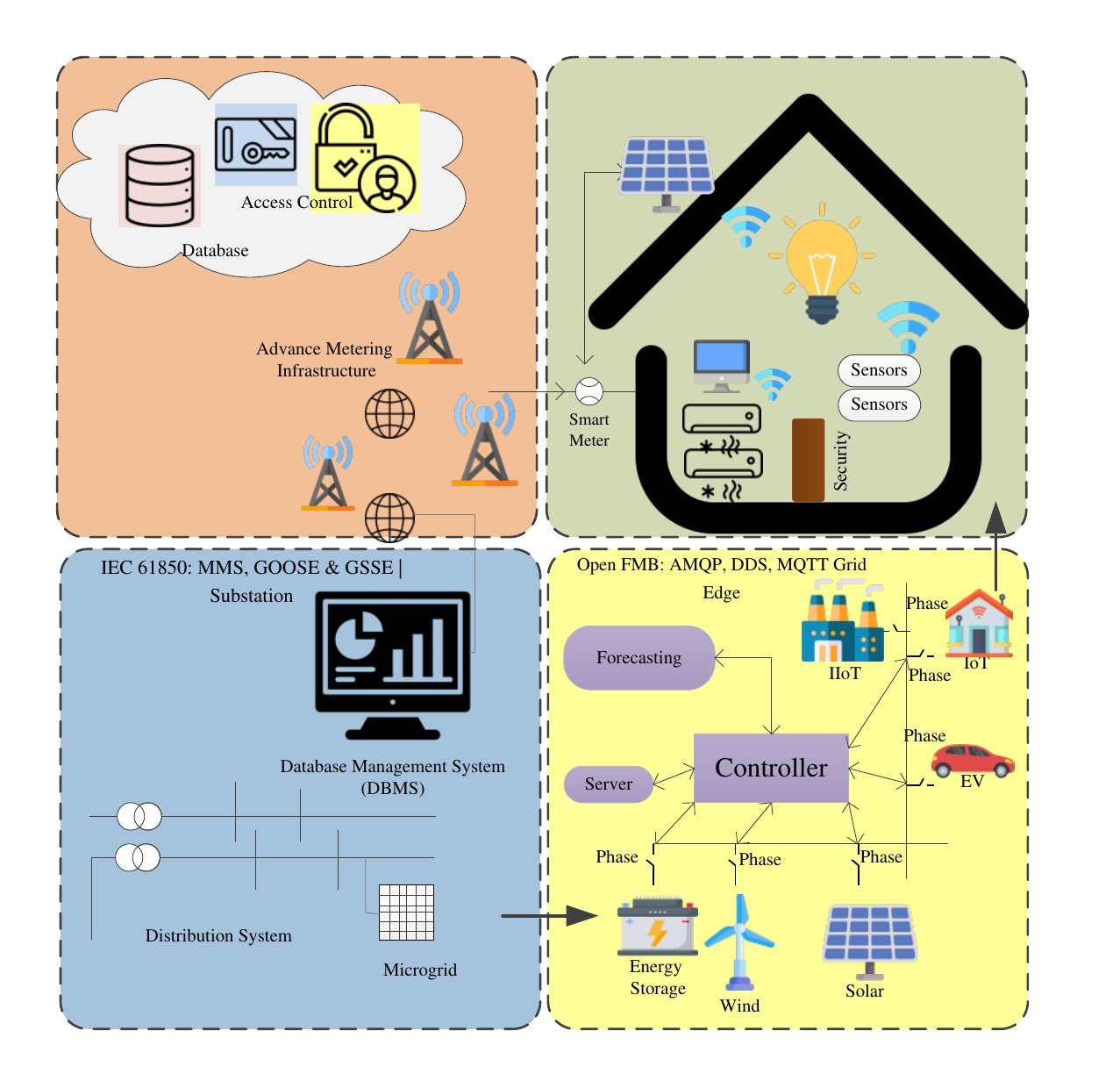}
    \vspace{- 0.3 cm}
  \caption{Application of IoT in smart distribution systems.}
  \label{smartgrid}
\end{figure}

From the perspective of the operator, DRRs are viewed as a virtual demand-side power plants. Nowadays, instead of dialing a costly unit during peak hours, operators are more likely to permit more reliable DRRs. After the emergence of structured power systems and the emergence of power markets, the demand response concept was developed. A change in power grids occurred with the restructuring and deregulation of the industry. In particular, the integration of the Advanced Metering Infrastructure (AMI), which is inspired by Internet of Things (IoT) technologies, has played a significant role in the modernization of today's power grid, including the collection of data that is necessary to manage load flows. Different level of IoT integration in smart cities in summarised form is shown in figure \ref{smartgrid}. 

\section{Smart Grids' Resiliency and Reliability}

In a power system, the system states are mostly determined using power flow methods. These methods, however, exclude non-linear variables and are generally used in the design phase of power systems, without considering real-time system states \cite{stott2009dc}. Other numerical techniques include graph theory, matrix principal, and some hybrid optimization techniques to handle the multilevel contingencies such as $N-k$ \cite{kaplunovich2016fast}. In addition to these techniques, ML techniques are also employed for the power flow calculations, which give really good approximation without involving any complex non-linear equations \cite{donon2020leap}. Furthermore, ML techniques can be compared to real-world problem-solving techniques and can also be used in practical scenarios due to their capability of providing real-time energy management and the ability to define the basic architecture for digital twin applications \cite{brosinsky2018recent}.

Due to the presence of different kinds of devices, tools, energy types, and their innate actions, variances in some factors in the energy field, and the unforeseen nature of some phenomena, it is imperative to exchange and analyze data in near real-time and to make decisions quickly. The needed activities must be carried out automatically, and the data must be swiftly and securely exchanged with the appropriate destinations. Hence, it is vital to integrate IoT technologies into components to benefit from networks' information technology. IoT-based devices can be sensors, meters, or controllers that have electric boards with microcontrollers and microprocessors that have the capability of sharing information. Additionally, it is necessary to maintain the accuracy and validity of information, which might be jeopardized by malicious cyberattacks or unintended disruptions.


\subsection{Resiliency}

\begin{figure}[h]
	\centering
    \includegraphics[width=10cm, height = 6 cm]{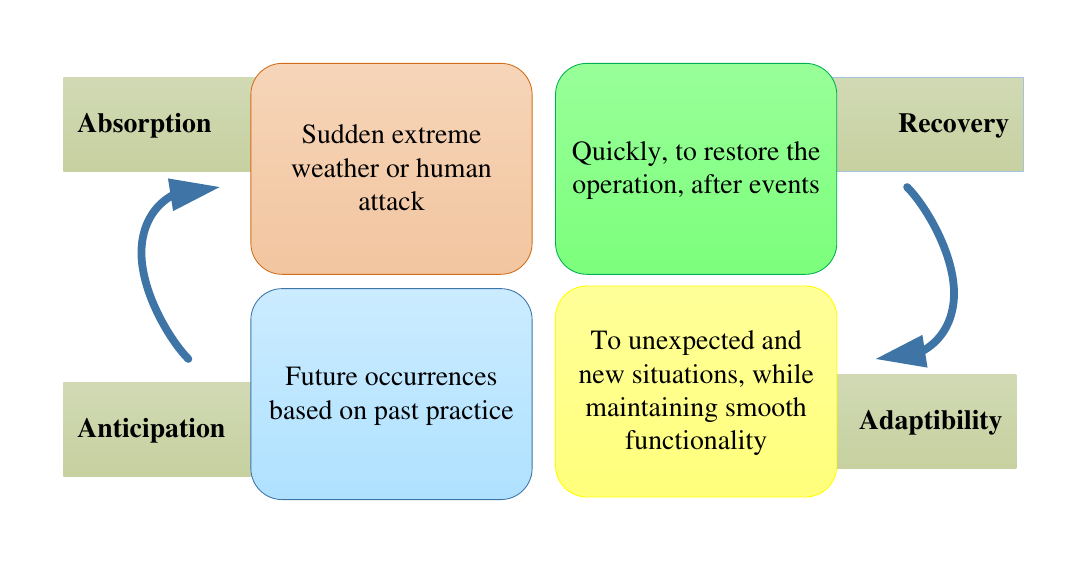}
	\caption{Resilient system.}
	\label{Resilient_system}
\end{figure}

Resiliency can be defined as the ability of the system to recover itself from any kind of disturbances \cite{mishra2021review}. 
From a power system perspective, the resilient system is the one, which has the capability of providing electricity even under extreme events, such as earthquakes, tornadoes, hurricanes, and cyber-physical attacks \cite{jena2020integrated}. Hence, the grid is considered to be resilient if it has four major properties, i.e., anticipation, absorption, recovery, and adaptability. Anticipation refers to the ability to avoid possible damages caused by natural disasters or unexpected events. Recovery is the ability of the grid to recover itself from damage that is least expected. Absorption is the capability of the grid to absorb the fluctuations and stress due to unexpected events, and adaptability is the property of a resilient grid to learn from previous events and adjust itself accordingly \cite{jufri2019state}, as shown in Fig.~\ref{Resilient_system}.  

\begin{figure}[h]
 	\centering
 	\includegraphics[width=10cm, height=8cm]{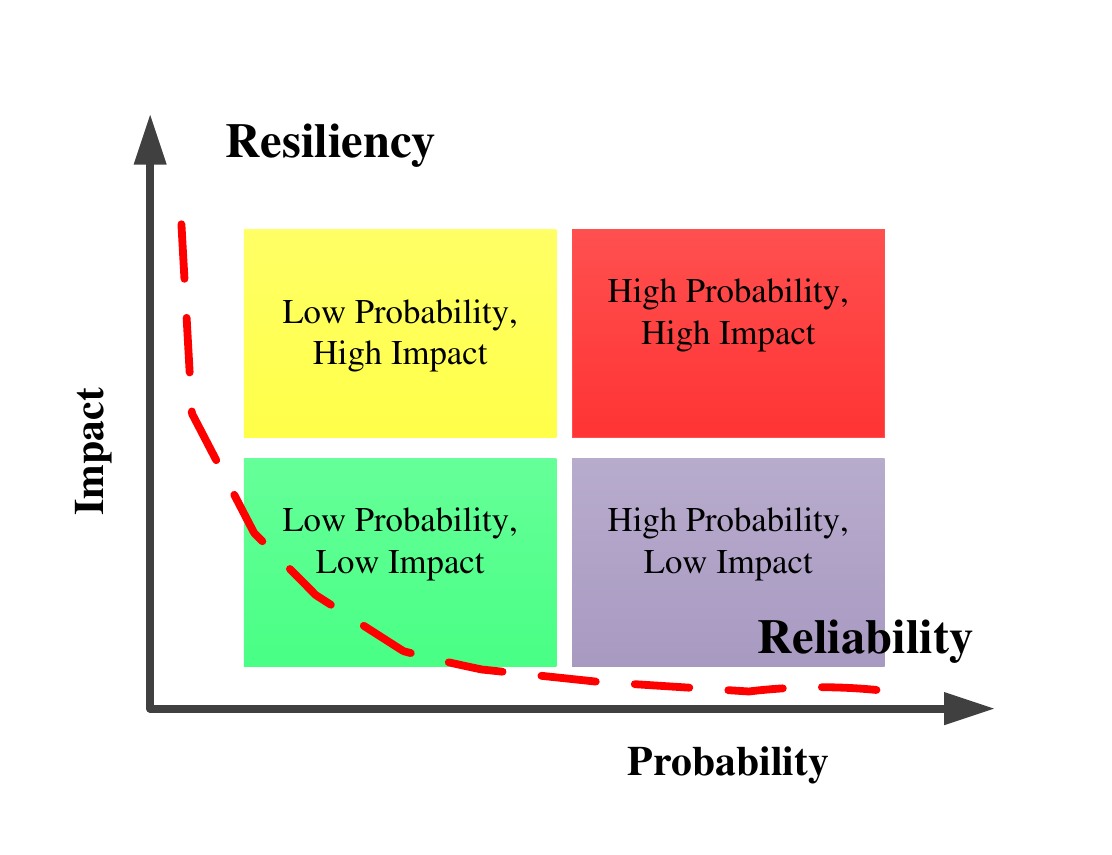}
 	\vspace{-0.7 cm}
 	\caption{Reliability vs resiliency \cite{mishra2021review}.}
 	\label{Reliability_resiliency}
 \end{figure}

Reliability and resiliency are the two important factors to evaluate the performance parameters of a power system. Some researchers treated them as the same entity, however, there are significant differences between them. The major distinguishing variables between reliability and resiliency are, low impact high probability (LIHP) and high impact low probability (HILP), respectively as shown in Fig.~\ref{Reliability_resiliency} \cite{mishra2021review}. 
The low probability and high impact terms describe events that are extremely rare, such as earthquakes and natural disasters, which can cause multilevel component failure that results in blackouts. The indices used to measure reliability are the system average interruption duration index (SAIDI) and the system average interruption frequency index (SAIFI). However, these terms exclude the power system parameters, which demonstrate the behavior of a system under unexpected events \cite{bie2017battling}. 
\subsection{Resiliency Characteristics and Metrices}
\begin{figure}[t]
 	\centering
 	\includegraphics[width=12cm, height=9cm]{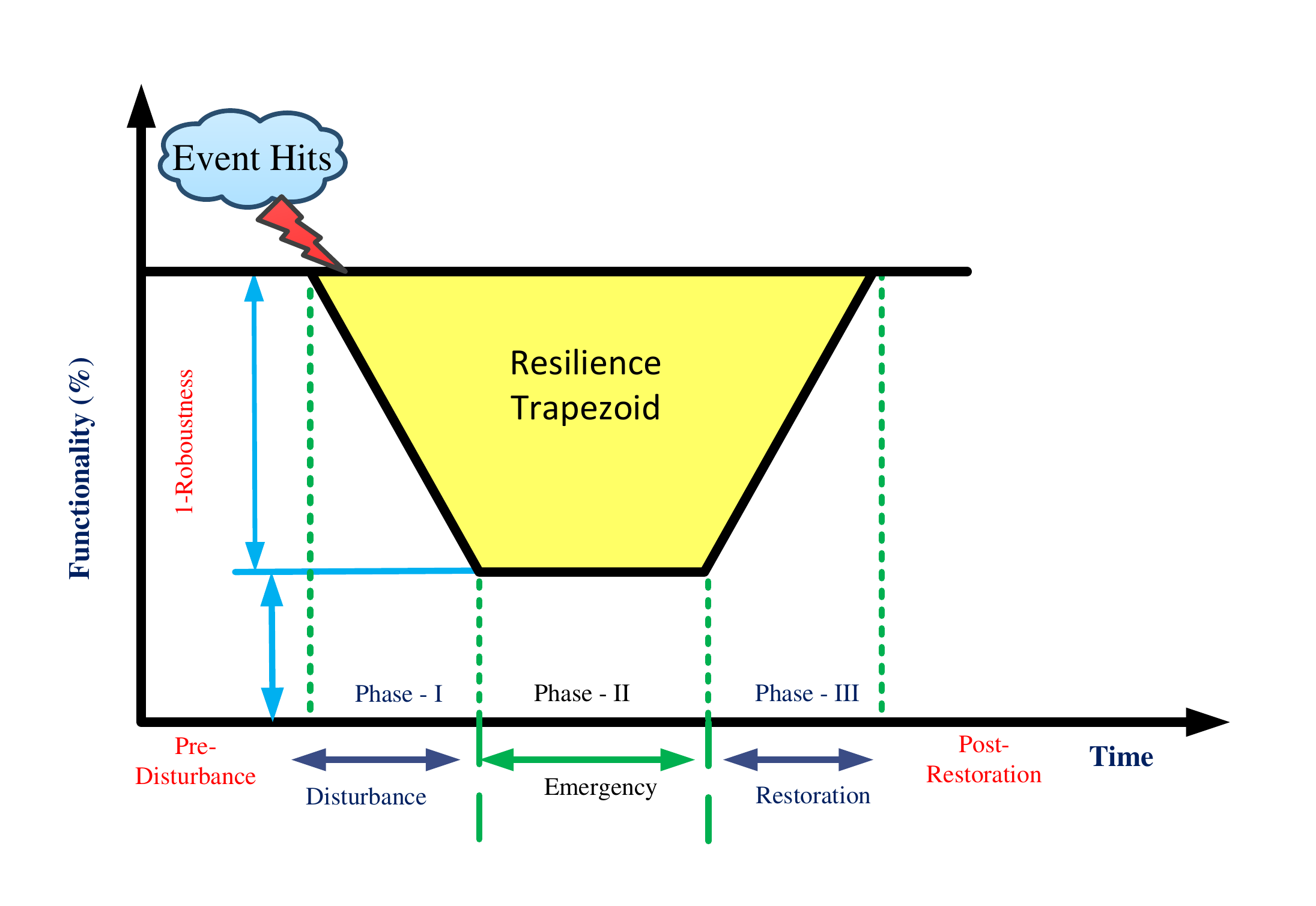}
 	\caption{Resilient trapezoid \cite{kroger2019achieving}.}
 	 	\label{trapezoid}
 \end{figure}

\begin{figure}[t]
 	\centering
 	\includegraphics[width=12cm, height=9cm]{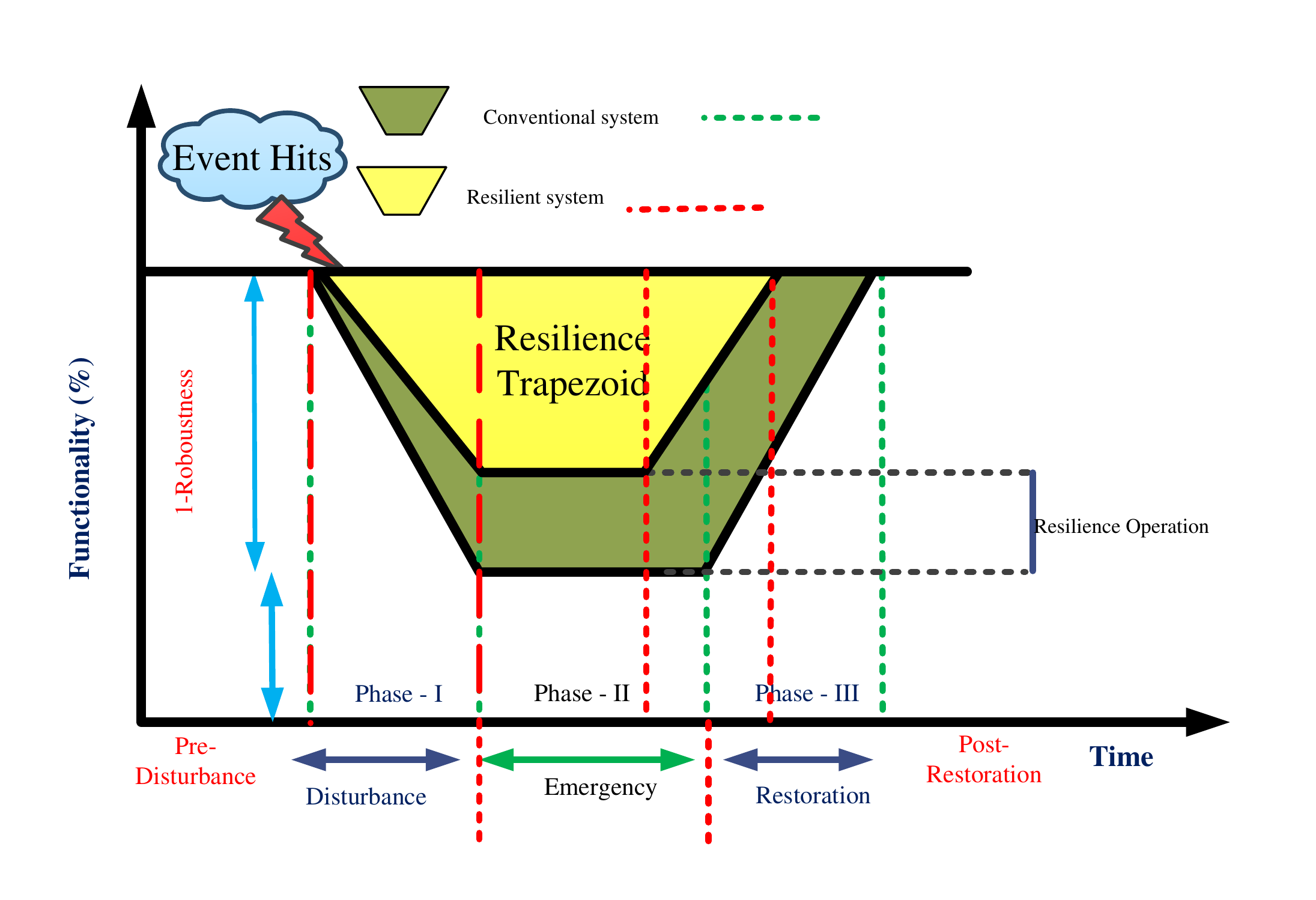}
 	\caption{Resilient vs. conventional system \cite{mishra2021review}.}
 		\label{Resilient_conventional}
 \end{figure}

The effectiveness and detailed analysis of the resilient methodology adopted for the cyber-physical system is determined using the resilience trapezoid \cite{kroger2019achieving}. The resilience trapezoid consists of three phases, as shown in Fig.~\ref{trapezoid}. In Phase I, unexpected events occur in the system that leads the system to instability. In Phase II, a protective mechanism is initiated to provide an emergency solution to the network. Finally, in Phase III, control actions take place based on the protective algorithms initiated in Phase II. The comparison between conventional and resilient systems is depicted in Fig.~\ref{Resilient_conventional}. It is worth noting that a resilient system significantly reduces the time taken to initiate the algorithm in Phase II and the stabilization time in Phase III. 

In order to evaluate the effectiveness of resiliency for the system, the resiliency metrics are most important. These include line trip probability (K), load not supplied probability (LNSP), load shedding probability (LSP), and an index representing grid recovery (G)\cite{shirzadi2018power}. The resiliency (\textit{r}) in terms of loss parameter can be represented as,

\begin{equation}
r=\frac{1}{loss},
\label{eq1}
\end{equation}

\begin{equation}
loss=\frac{SP_o-SP\textsubscript{min}}{SP\textsubscript{min}},
\label{eq2}
\end{equation}

\begin{equation}
loss=\int_{t_1}^{t_4}\frac{SP_o-SP\textsubscript{min}}{SP\textsubscript{min}} dt,
\label{eq3}
\end{equation}

\begin{equation}
loss=\frac{1}{t_4-t_1}\int_{t_1}^{t_4}\frac{SP_o-SP\textsubscript{min}}{SP\textsubscript{min}} dt.
\label{eq4}
\end{equation}
Where $SP$ is the performance of the system under an unexpected event, $loss$ is the unused resources and $t$ is the time.
 
 \section{Resilient strategies}
The resiliency strategies of a power system network usually fall into two categories, i.e., soft and hard resiliency also known as system hardening. System hardening usually refers to the physical changes in the infrastructure, which require a lot of funding and time. Hence, achieving system hardening at a reduced overall cost is the need of time. System hardening is achieved by laying underground cables for the transmission and distribution system, relocating the substation to a safer side, providing alternate routes for power transfer in case of line damage, and increasing the strength of the poles. 

The implementation of smart grids opens a new area to design a resilient power system network. Through advanced electronic components, the fault location can be identified faster and can provide more robust operational strategies to enhance the system resiliency, such as real-time energy and risk management, disaster prediction and damage estimation, demand and supply management, island operation of the microgrid, and control and operation of RERs \cite{ali2019optimum}. 

Recent related works on system hardening and optimization algorithms to improve the resiliency of the system is listed in Table~\ref{Optimization_resilience}. A major drawback associated with these techniques is their inability to handle highly non-linear parameters. Moreover, these optimization techniques require complex mathematical equations, which increase the computational cost and take a long time to find optimal states.

\begin{table*}[t!]
	\caption{Optimization-based resilience techniques.}
	\begin{tabular}{|p{0.8in}|p{1in} |p{1in}| p{1.5in}| p{1.5in}|}
	    \hline
		Area of study& Model&Objective function& Pros& Cons\\ \hline
		
        System hardening \cite{wang2017robust} & Iterative defender against defender model (DAD)  & Optimizing the cost associated with system hardening & Providing optimal strategy for system hardening against N-k events.  & Only portion of the network is hardened, which makes the rest vulnerable. \\ \hline
        
        Coordination planning \cite{sedzro2017allocation} & Mixed integer linear programming (MILP)  & Identifying critical load& Increasing the response time of microgrid to quickly restore power under unexpected event.   & Presented short-term planning, with minimum constraints. \\ \hline
        
        Coordination planning \cite{lee2019energy}&MILP&Load shedding reduction & Optimal utilizing the reserve resources to meet the demand in disaster.& Neglecting the constraint will deviate the solution from its optimal state value. \\ \hline
        
        Planning resources \cite{arab2015stochastic}&Stochastic linear programming  & Cost minimization & Designing an algorithm for optimal unit commitment in order to provide power with less cost.& Does not provide real-time energy management in terms of demand and response. \\ \hline
    
    Power restoration \cite{yao2018transportable}&MILP&Minimizing the operational cost& Providing the cost balance between fixed energy resources and reserve in a microgrid.& The constraints related to transmission networks are neglected. \\ \hline
    
    Power reconfiguration \cite{jin2018planning}& Moment-based design  & Cost and energy optimization&Developing the probabilistic model to effectively integrate the RERs on the distribution side to reduce restoration time of power.r& The transients due to RERs are neglected. \\ \hline
    
    Repairing of the network \cite{chen2018toward}&Synthetic model & Energy and crew management&Formulating the synthetic model to repair the distribution network for a crew member.&The repair process is done in steps, so it takes time to repair the system in each zone.\\ \hline
	\end{tabular}
	\label{Optimization_resilience}
\end{table*} 

\subsection{Data Driven Techniques to Improve Resiliency}

In this subsection, a comparative analysis of different ML techniques is performed, which can be utilized for an efficient transformation of the smart grid environment into smart cities.

\subsubsection{Model predictive base power outage management}

In \cite{farzin2016enhancing}, the authors developed a hierarchical outage management scheme to enhance the resiliency of a power system by routing unused power to an interconnected microgrid during disaster events. The model predictive control (MPC) is designed to schedule the available energy resources. The distribution operator then utilizes those available energy resources in single microgrid and routes it to that microgrid where power is needed in order to balance the load. The main aim is to minimize the overall cost, as shown in the below objective function.

\begin{equation}
Min\sum_{t=1}^{n}\bigtriangleup{t}\Big\{\sum_{1}^{nG_i}cs_g\textsuperscript{G,i}P\textsubscript{g,t}^i+\sum_{1}^{nL_i}cs_l\textsuperscript{L,i}CL\textsubscript{l,t}^i\\+\sum_{1}^{nR_i}cs_r\textsuperscript{R,i}P\textsubscript{r,t}^i+\sum_{1}^{nB_i}cs_b\textsuperscript{B,i}P\textsubscript{b,t}\textsuperscript{i,dis}\Big\}.
\end{equation}

Where, $cs_g$, $cs_l$, $cs_r$ and $cs_b$ are cost associated with generation sources, load, RERs and storage devices respectively, $P$ stands for active power and $CL$ stands for load curtailment. The $nG$, $nL$, $nR$, and $n_B$ stand for the number of generation, the total number of loads, RERs, and storage devices, respectively. The total power generated by conventional and individual RERs must be equal to the net demand.

\begin{multline}
\sum_{1}^{nG_i}P\textsubscript{g,t}^i+\sum_{1}^{nR_i}P\textsubscript{r,t}^i+\sum_{1}^{nB}(P\textsubscript{b,t}\textsuperscript{i,dis}-P\textsubscript{b,t}\textsuperscript{i,chr})=\sum_{1}^{nL_i}(Ld\textsubscript{l,t}^i-CL\textsubscript{l,t}^i).
\end{multline}

The power generation constraints will be,
\begin{equation}
P\textsuperscript{min}_g\leq P\textsubscript{g,t}^i\leq P\textsuperscript{max}_g,
\label{eq7}
\end{equation}

\begin{equation}
 P\textsubscript{g,t}^i- P\textsubscript{g,t-1}^i\leq RU_g,
 \label{eq8}
\end{equation}

\begin{equation}
P\textsubscript{g,t-1}^-P\textsubscript{g,t}^i- i\leq RD_g.
\label{eq9}
\end{equation}

 Where, (8) and (9) show ramp-up and ramp-down constraints. The RERs and load constraints are as follows,
 \begin{equation}
 0\leq P\textsubscript{r,t}^i\leq P\textsubscript{r,t}\textsuperscript{max},
 \label{eq10}
 \end{equation}
 
 \begin{equation}
 0\leq CL\textsubscript{l,t}^i\leq Ld\textsuperscript{l,t}_i.
 \label{eq11}
 \end{equation}

\subsubsection{Logistic regression and support vector machine (SVM) based resiliency}

The regression-based analysis for identifying component failures during hurricanes was presented in \cite{eskandarpour2016machine}. Logistic regression is performed to find the decision boundary between healthy and failed components. The second order polynomial equation is used for fitting, which is,
\begin{equation}
h(x,k)=k_o+k_1x_1+k_2x_2+k_3x_1^2+k_4x_2^2+k_5x_1x_2.
\label{eq12}
\end{equation}

Where, $k$ is the characteristic parameter and Sigmoid as an activation function is used, which can be represented as,
\begin{equation}
F(x,k)=\frac{1}{1+e\textsuperscript{-h(x,k)}}.
\label{eq13}
\end{equation}

The main aim is to minimize the cost function during training and testing processes, where, $\lambda$ is the regularization parameter.
\begin{equation}
H(x,k)=\frac{1}{m}\sum_{i=i}^{n}c(F(x,k),y)+\frac{\lambda}{2m}\sum_{i=1}^{5}k_i^2.
\label{eq14}
\end{equation}

However, the simple logistic regression model is unable to give results as it requires large datasets for training purposes. To overcome this issue, the authors in \cite{maharjan2019machine} performed a classification of the failure of the component through the SVM algorithm and found that it gave very good results by simply modifying the cost function.

\subsubsection{Long short term memory (LSTM) based attack detection to improve resiliency} 
The LSTM is an improved version of recurrent neural networks. The importance of LSTM networks is that they can provide very significant results for sequential problems \cite{wang2019probabilistic}. The power system is a time series-based system and attacks that happen on the network are also time-dependent. These attacks can corrupt the measurements of sensors like phasor measurement units (PMUs) installed in the network at different locations. The PMUs mostly display voltage and phase angle for the system network. If these two states are corrupted, it will lead the system to instability. The LSTM algorithm can be used to detect those corrupt states and will help in stabilizing the system. For state prediction, LSTM is adopted due to its ability to model time series data more accurately since it has a memory cell, which stores previous hidden states and utilizes it for the training of LSTM. The LSTM working consists of the following steps:

Prediction Step-1: This step is also called the forget step. In this step, LSTM will decide what information is unnecessary and needs to be removed from the memory cell. It  decides that by using the following equation.
\begin{equation}
v_t=\sigma(We_v.[hs\textsubscript{t-1},x_t]+bi_v)
\label{eq15}
\end{equation}  

Prediction Step-2: This step is called the cell update step. As shown in equations ~\ref{eq16}, ~\ref{eq17}, and ~\ref{eq18}, the sigmoid layer $\sigma$ within a single LSTM cell will decide what new values will be added and the $tanh$ layer will create a new vector for the proposed cell state $\tilde{C}_t$, which only depends on the current input $x_t$t and hidden state $hs\textsubscript{t-1}$, and will be added to the old cell state $C\textsubscript{t-1}$.

\begin{equation}
in_t=\sigma(We\textsubscript{in}.[hs\textsubscript{t-1},x_t]+bi\textsubscript{in}),
\label{eq16}
\end{equation}

\begin{equation}
\tilde{C}_t=tanh(We_c.[hs\textsubscript{t-1},x_t]+bi_c),
\label{eq17}
\end{equation}

\begin{equation}
C_t=v_t*C\textsubscript{t-1}+in_t*\tilde{C}_t.
\label{eq18}
\end{equation}

Prediction Step-3: At this step, the filtered output will be displayed. First, $hs\textsubscript{t-1}$ and $x_t$ are passed through the sigmoid layer and the current cell state $C_t$ is passed through $tanh$ layer so that the final output $hs_t$ will be found for a single LSTM cell at the current time instant.

\begin{equation}
ou_t=\sigma(We\textsubscript{ou}.[hs\textsubscript{t-1},x_t]+bi\textsubscript{ou}),
\label{eq19}
\end{equation}

\begin{equation}
hs_t=ou_t*tanh(C_t),
\label{eq20}
\end{equation}

\begin{equation}
Er=\frac{1}{2}\sum_{1}^{n}(actual-hs_t)^2.
\label{eq21}
\end{equation}

Where, $We_v$, $We\textsubscript{in}$, $We\textsubscript{c}$, $We\textsubscript{ou}$, $bi\textsubscript{v}$ $bi\textsubscript{in}$, $bi\textsubscript{c}$, $bi\textsubscript{ou}$ are weights and biases of neural network layer. The  mean square error $Er$ is computed using (21). At the beginning, $hs_t$ and $C_t$ are initialized with zero vectors and then network weights and the matrix vectors are updated using back propagation through time algorithm (BPTT). A brief review of some advanced data driven ML techniques are listed in Table.~\ref{Data_driven_ML}.

\begin{table*}
	\centering
	\caption{Data driven ML-based resilience enhancement.}
	\begin{tabular}{|p{0.5in}|p{1.8in}|p{1.8in}|p{1.8in}|}
		\hline
		Year&Technique  &Pros &Cons  \\ \hline
	    2012\cite{verma2012supervised}	& Deep feed-forward neural network (DFFNN)  & The supervised DFFNN is proposed to enhance system security and detection of contingency analysis. & Does not provide control action when unexpected events take place in the network.  \\ \hline
	    2018\cite{donnot2018anticipating}	& DFFNN with dropout algorithm  &The algorithm is used for the topology search in transmission network under faults. & The network is tested under N-1 and N-1-1 contingencies while neglecting N-k contingency.  \\ \hline
		2019\cite{kim2019graph}& Graph convolutional neural network (GCNN)  & GCNN is used to observe the available energy resources in the system and perform estimation for optimal load shedding. & The load shedding cause revenue loss. \\ \hline
		2019\cite{donon2019graph}& Graph neural network & The technique is used to predict the power flow in the network. & The model is used for DC power flow while neglecting unexpected events. \\ \hline
		\end{tabular}
		\label{Data_driven_ML}
\end{table*}

\begin{figure}
         \centering
         \includegraphics[width=12cm, height=9cm]{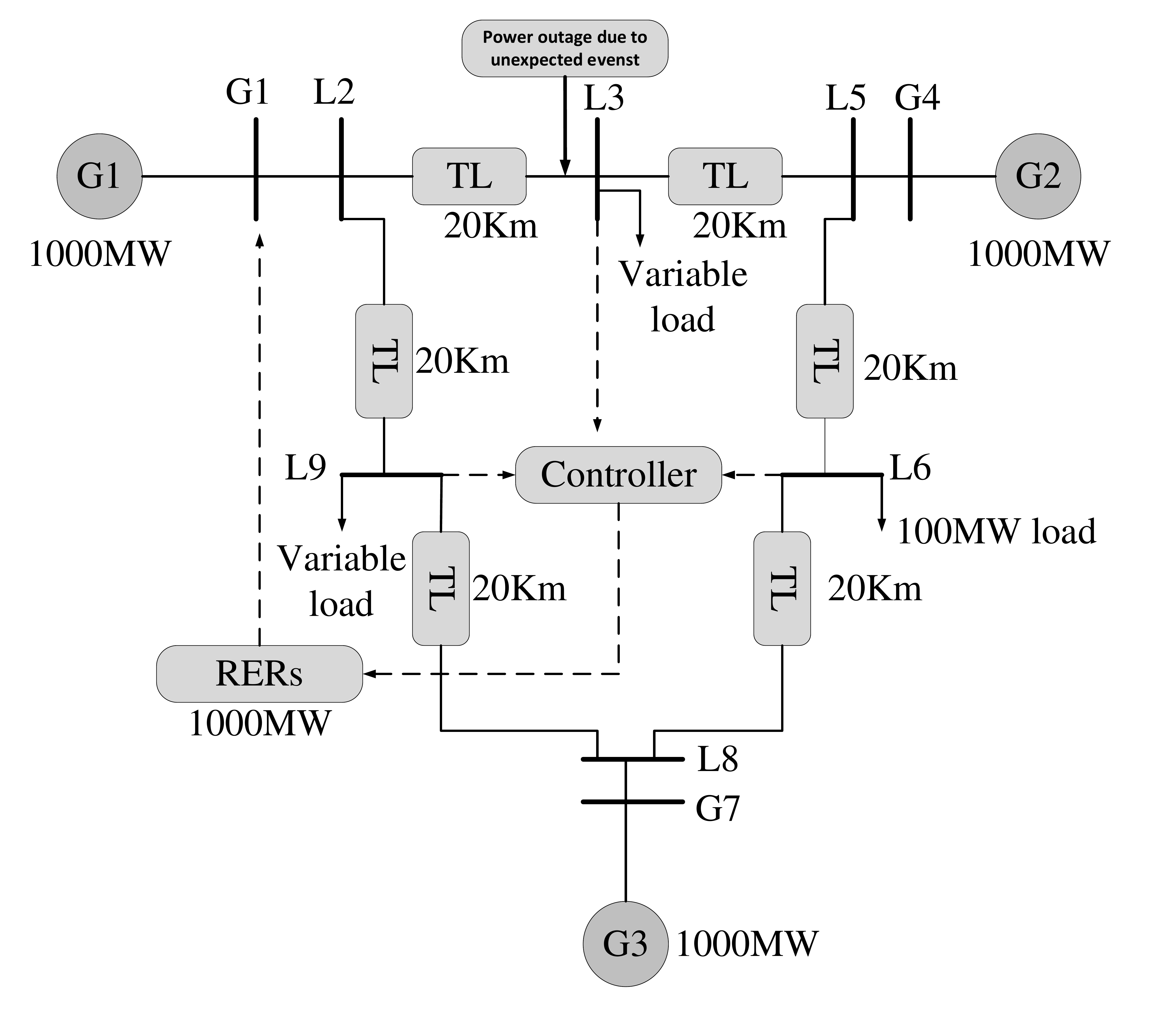}
         \caption{IEEE 9bus system.}
         \label{9bus}
\end{figure}

\begin{figure}[h]
         \centering
         \includegraphics[width=12cm, height=9cm]{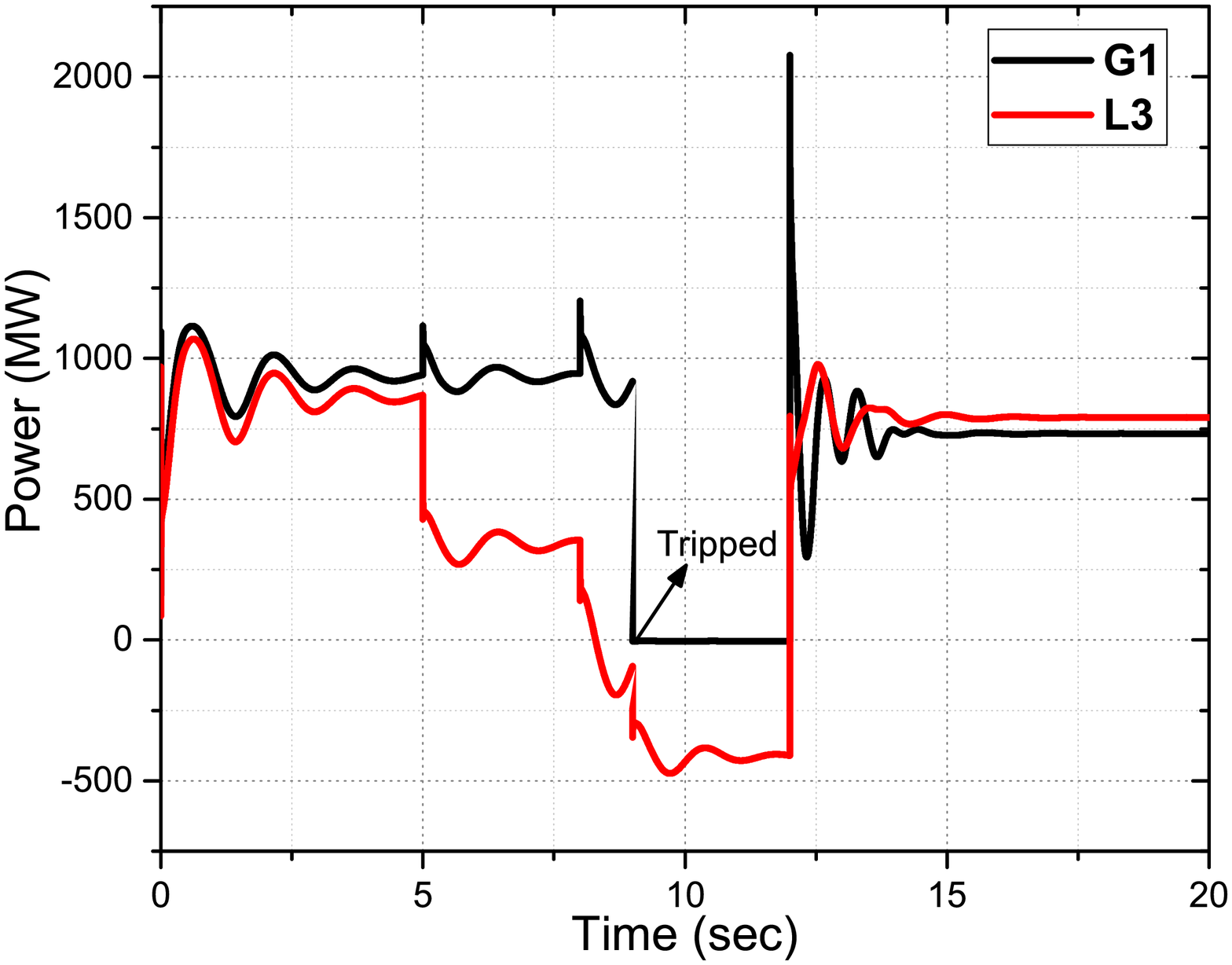}
         \caption{N-k contingencies line G1 disconnected.}
         \label{N_k_disastersa}
\end{figure}

\begin{figure}
         \centering
         \includegraphics[width=12cm, height=9cm]{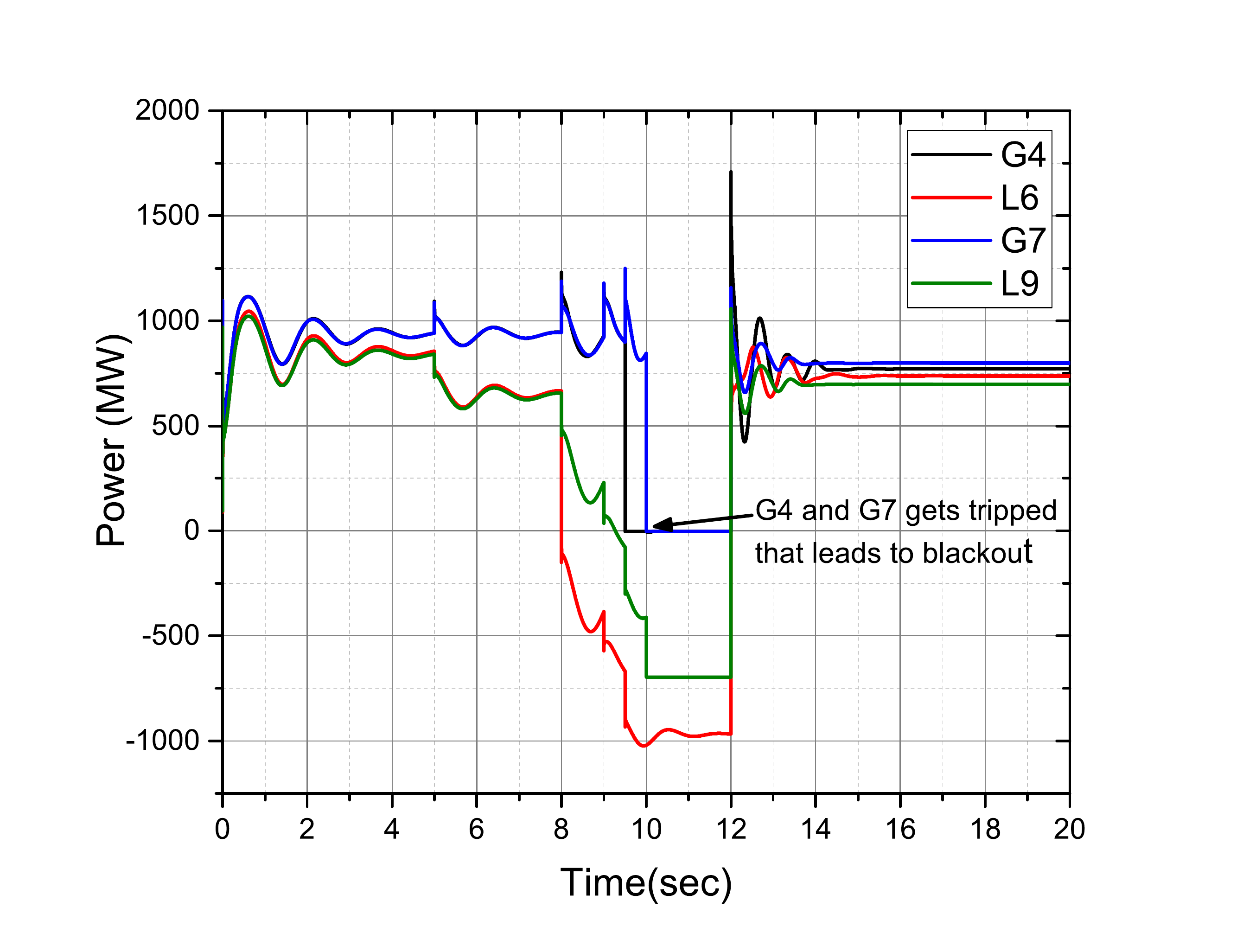}
         \caption{N-k contingencies line G4 and line G7 disconnected.}
         \label{N_k_disastersb}
     \end{figure}
\begin{figure}
	\centering
	\includegraphics[width=12cm, height=9cm]{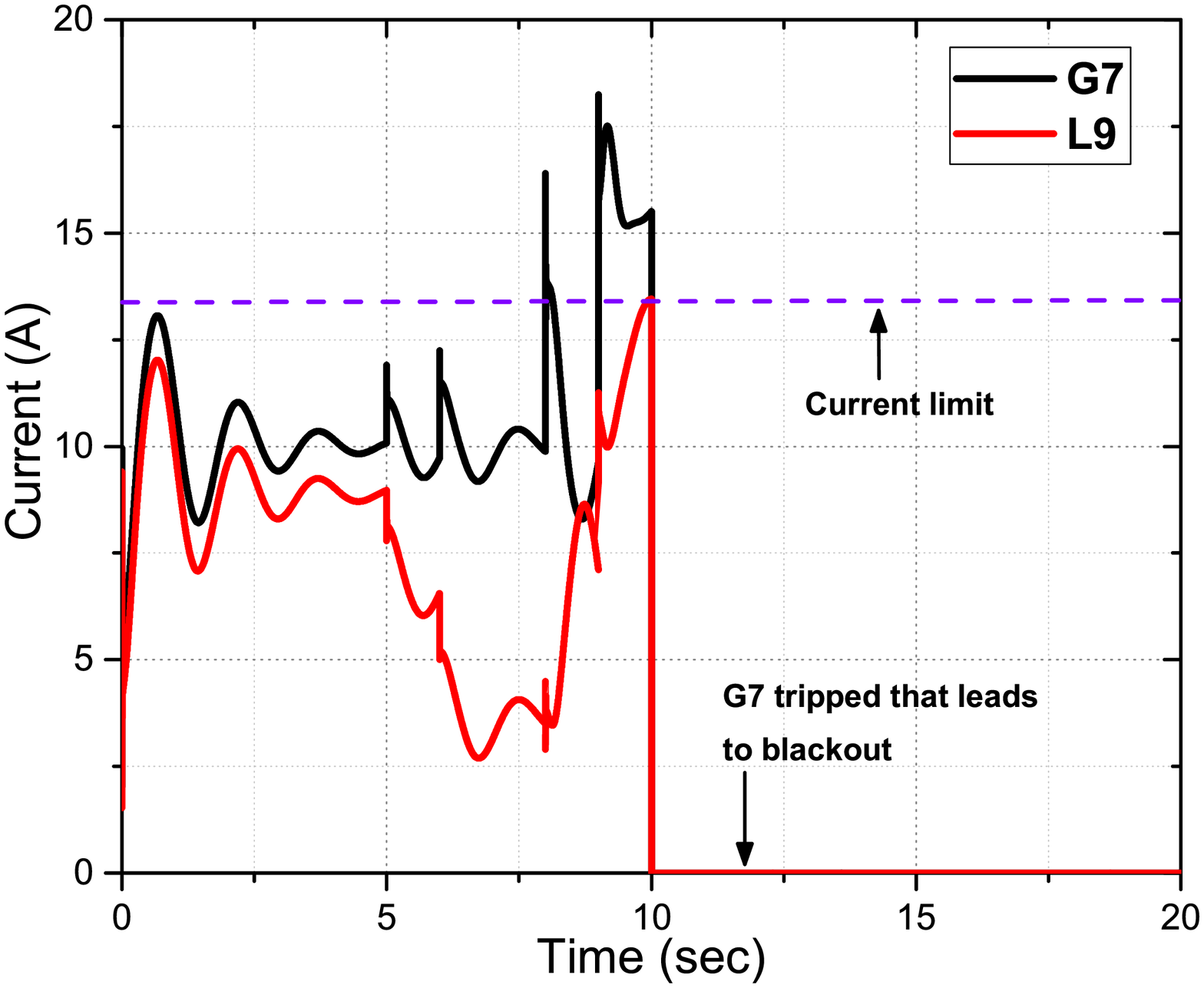}%
	\caption{N-k contingencies current profile at lines G7 and L9.}%
	\label{Power_loss}%
\end{figure}
\section{Comparative Analysis of ML Techniques Using Analytical Reasoning}
\subsection{N-k Contingencies}

The IEEE 9 bus system under N-k contingency is shown in Figure~\ref{9bus}, where $G$ corresponds to generation and $TL$ and $L$ correspond to transmission lines and loads, respectively. As an unexpected event occurred at time 8 seconds, line G1 which supplies power to the network got disconnected, as shown in Fig.~\ref{N_k_disastersa}. In smart grids, a network is interconnected and the load is distributed among different lines to reduce stress on individual lines, which is why the power in line G4 was rerouted to stabilize the system. However, the demand was higher than nominal and as a result, line G4 also got disconnected at 9.5 seconds. Furthermore, more negative power was accumulated on line L6 as shown in Fig.~\ref{N_k_disastersb}. Now, all the demand was shifted to the last line G7, and more power deficiency was observed on line L9. Line G7 was unable to handle the situation, so it also got disconnected at 10 seconds and a complete blackout occurred in the network from 9 seconds to 12 seconds, as shown in Fig.~\ref{N_k_disastersb}. At 12 seconds, when the disturbance was cleared, the system stabilized within 3 seconds, but the blackout was still present from 9 seconds to 12 seconds. The variation in power profile after 12 seconds was all due to the unbalance demand distribution. 

Figure ~\ref{Power_loss}  16 shows the current profile at respective buses during the N-K contingency. The maximum current limits for the transmission lines are at 13 A. At 5 seconds, the current profile at line L3 overshoot from the limit; as a result, line G1 tripped at 6 seconds.
The power system network was interconnected, so line L3 is supported by lines G4 and G7 due to which current residuals were presented.
The load is supported up to some instants and at 8 seconds, the current at line L6 also crosses the limit, in which line G4 trips at 9 seconds.
In a similar way, line G7 also trips at 10 seconds which leads to a complete blackout of the power system from 10 seconds onward, as shown in Fig.~\ref{Power_loss}.

\subsection{Comparison of SVM, Logistic Regression and LSTM in Stabilizing the System Under Disturbances}

\begin{figure}
	\centering
	\includegraphics[width=12cm, height=9cm]{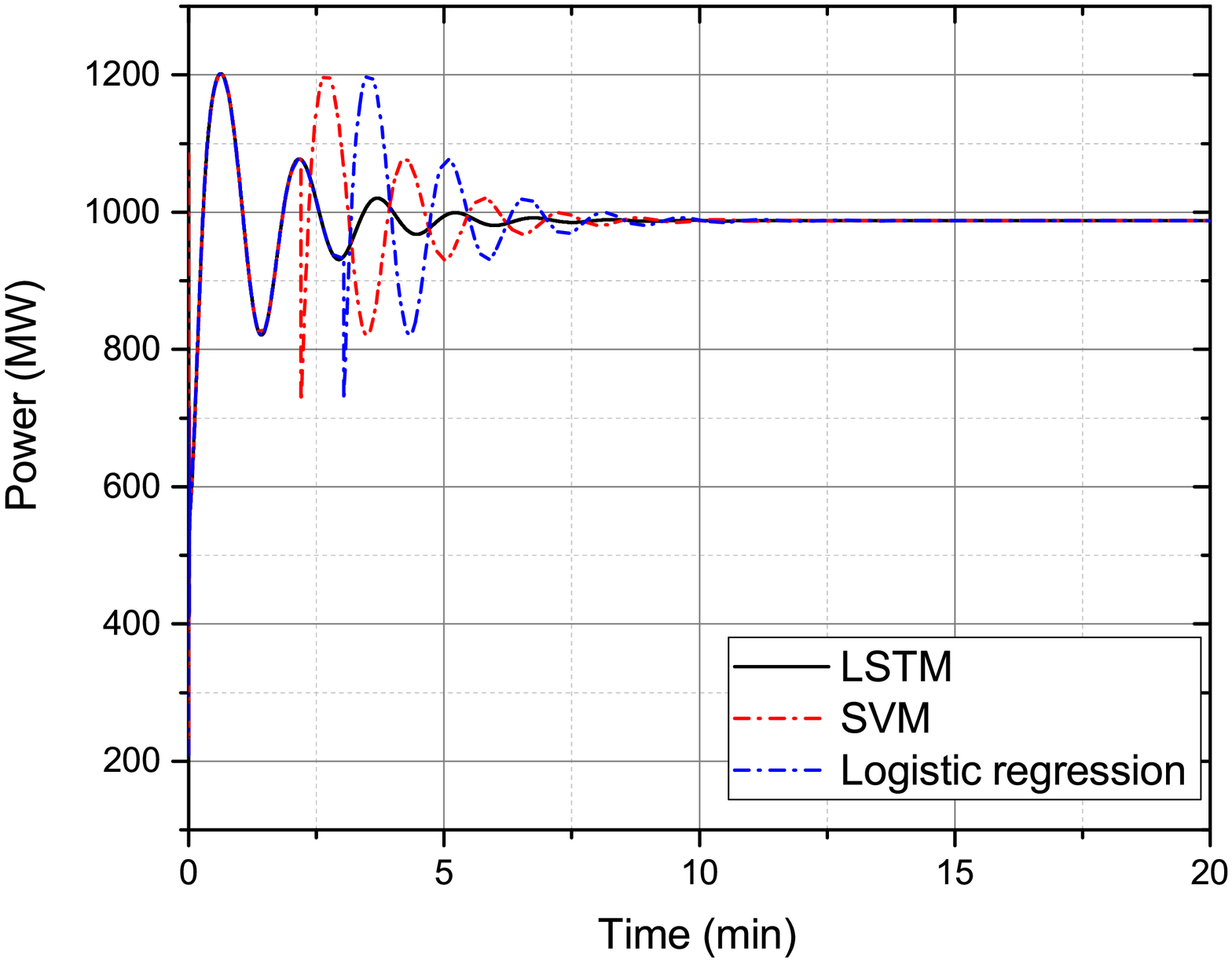} 
	\caption{System stabilization under disturbances.}%
	\label{System_stabilization2}%
\end{figure}

\begin{figure}
	\centering
    \includegraphics[width=12cm, height=9cm]{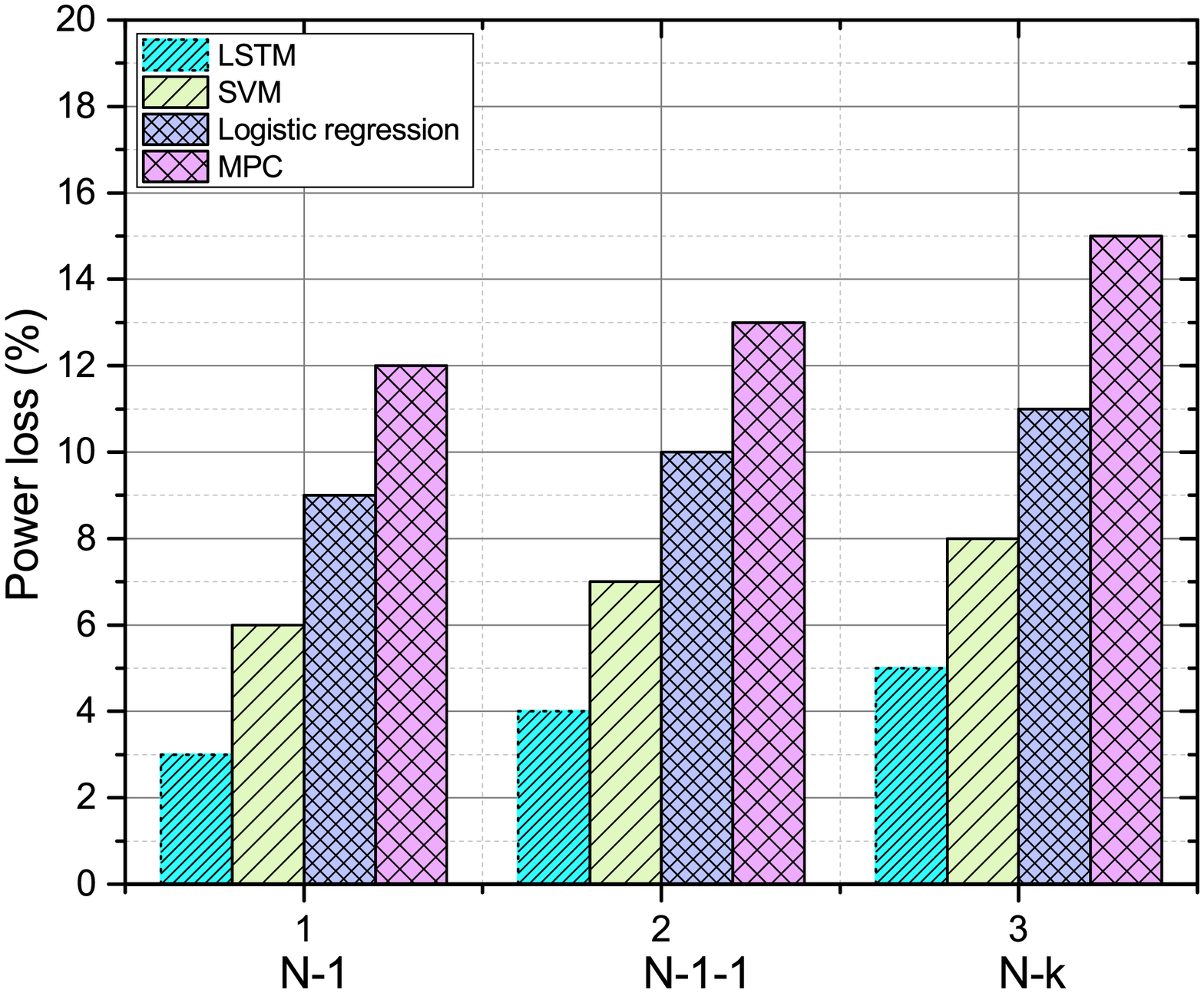}%
    \caption{Power loss comparison under N-1, N-1-1, and N-k contingency.}
    \label{Power_loss2}%
\end{figure}

Unexpected events in the network can cause severe conditions in the network. Based on the LSTM, SVM, and logistic regression, these fluctuations in the network can be detected and a control action taken. It can be visualized in Fig.~\ref{System_stabilization2} shows that incorporating RERs into the control action based on the LSTM algorithm stabilizes the system within 5 minutes compared to 7 to 8 minutes if the control action is taken based on the prediction provided by SVM and regression. This shows that LSTM-based prediction provides good approximation in less time. To further illustrate the effectiveness of the LSTM, the percentage power loss function is shown in Fig.~\ref{Power_loss2}, where it can be observed that LSTM-based estimations reduce the power loss under unexpected events that are N-1, N-1-1 and N-k contingencies compared to other state of art techniques such as SVM, logistic regression and MPC. 


\section{Research Challenges}

\subsection{Fault Tolerance}

The likelihood of several hardware components and various data sources failing in a disaster is high owing to physical damage, depleted batteries, and communication channel failure. Therefore, to sustain system availability in a DRSC context, sources of information should be able to deliver data even during blackouts and infrastructure degradation. The construction of a backup power consumption mechanism and an alternative communication channel is essential. Additionally, the environment must-have features like a cloud-based storage system with distributed computing capability that may be used in the event that the primary system fails.

\subsection{Meta Data}

In order to identify and manage the data sets for a time and data-sensitive application, metadata is crucially significant. Moreover, a variety of integration-related problems and issues about data quality may be resolved by metadata, allowing for the presentation of real datasets for study. In the context of the DRSC environment, the primary metadata aspects that need to be identified are the source of information, material, geographical references and time. However, the gathering and administration  of metadata for diverse large data sources, particularly during disasters, is a significant challenge. In addition, since there are so many different data streams and data formats, creating and preserving metadata in the big data model is especially tricky. While some data may have some form of metadata associated to them already, most do not. Lastly, datasets can become highly complicated as multiple sensors are used by public and commercial groups for a variety of objectives.

\subsection{Privacy and Security}

As BDA primarily uses personal information to provide the desired results, privacy concerns have been a significant issue. Concerns regarding probing, segregation, theft, and monitoring are rising as a result of the public examination of personal information \cite{tene2011privacy}.

Social media databases, for instance, provide user location and personal information that might be utilized by hostile parties for bad intentions, particularly during emergencies like civil conflicts. IoT's utilization and productivity are constrained by a variety of security and privacy concerns that are experienced by its end users \cite{Hameed2019}. Additionally, the Hadoop Ecosystem lacks sufficient security solutions for a lot of technologies \cite{Kim2014}. Moreover, with the accessibility of enormous and extensively detailed data, the apparent or impending security threat can seriously undermine public confidence in data aggregation and exchange \cite{Sicari2015}.  In order to ensure the integrity, accessibility, and privacy of disaster-related information, and to prevent its misuse, appropriate security measures and authorization controls are crucial. 

\subsection{Time Constraints}

In crisis management, time is of the essence since an immediate response will save lives. However, it might be challenging to use vast amounts of diverse data to quickly extract desired conclusions for emergency actions. Even with sophisticated technologies, determining the proper approach alone requires time-consuming and complicated steps that include data gathering and standardization. Furthermore, unstructured data might exacerbate the issue by necessitating various implementation techniques based on the type. The generation of high-quality information from enormous amounts of heterogeneous data in a certain length of time is a significant problem for the currently available approaches and technologies.

\subsection{Standardization}

The adoption of technical and administrative advancements, the endorsement of system effectiveness, and the provision of reliable application, regulation, and future study guidelines are all facilitated by standards. There is a great need and opportunity for communication, information management standards, and security requirements to be re-assessed with the expanding use of BDA and IoT technologies. Setting and adhering to standards for various emerging technologies while taking into account the need for disaster management to receive correct responses in almost real-time is highly difficult.

\subsection{Grid resiliency}

Some research challenges to enhance the grid's resiliency are as follows:

\begin{itemize}
	\item Due to various limitations in the existing resiliency techniques, more robust techniques will be required for quick recovery of the system under disturbances.
	\item Incorporation of RERs causes fluctuations in the power profile, which leads the system to instability. Thus, more advanced and sophisticated techniques are required to enhance the resiliency of renewable integrated grids.
	\item Unexpected events can distort the communication between the sensor networks and the control center. Moreover, communication networks are also vulnerable to cyber threats. Cost-effective system hardening and resiliency against cyber threats will be needed. 
	\item Multi-event assessment by adopting deep reinforcement learning and real-time energy management using intelligent and adaptive control strategies under disturbances needs to be investigated. 
\end{itemize}

\section{Conclusion}
Resilient smart city infrastructure is a need of today's society. In this regard, this paper effectively discussed how to achieve resilient infrastructure using advanced techniques. For instance, how BDA can be utilized for designing and monitoring the infrastructure in order to make it resilient against disasters and man-made events like cyber attacks. Moreover, the need for efficient operation of smart drones for monitoring activities and data collection in smart cities has also been discussed. The paper has presented a holistic view of a resilient and sustainable smart city architecture that has utilized IoT, BDA, smart drones, and smart grids through the intelligent integration of RERs. Through a case study, the impact of disasters on the power system infrastructure in smart cities
has also been investigated and different types of optimization techniques that can be used to sustain the power flow in the network during disturbances are compared and analyzed. Lastly, a comparative review analysis of different data-driven ML techniques for sustainable smart cities has been performed along with a discussion on open research issues and challenges.

 \bibliography{cas-refs}





\end{document}